\documentclass[10pt]{article}

\usepackage{comment}
\usepackage{amsfonts}
\usepackage{amssymb}
\usepackage{pgfplots}
\usepackage{algorithm}
\usepackage{algorithmic}   
\usepackage{authblk}
\usepackage[margin=1in]{geometry}

\input{Qcircuit}

\newcommand{\braket}[2]{\left<#1|#2\right>}

\newcommand{\nat} {{\mathbb N}}

\newcommand{\reals}{{\mathbb R}}

\newtheorem{theorem}{Theorem}
\newtheorem{rem}{Remark}

\newcommand\e{\varepsilon}

\newcommand{\norm}[1]{\| #1\|}

\title{Approximating Ground and Excited State Energies on a Quantum Computer}
\author{Stuart Hadfield, Anargyros Papageorgiou} 
\affil{Department of Computer Science,\\ Columbia University}

\begin{document}
\maketitle

\begin{abstract}
Approximating ground and a fixed number of excited state energies, or equivalently low order Hamiltonian eigenvalues, is an important but computationally hard problem. 
Typically, the cost of classical deterministic algorithms grows exponentially with the number of degrees of freedom. Under general conditions, and using a perturbation approach, we provide a quantum algorithm that produces estimates of a constant number $j$ of different low order eigenvalues. The algorithm relies on a set of trial eigenvectors, whose construction depends on the particular Hamiltonian properties. We illustrate our results by considering a special case of the time-independent Schr\"odinger equation  with $d$ degrees of freedom. Our algorithm computes  estimates of a constant number $j$ of different low order eigenvalues with error $O(\e)$ and success probability at least $\frac34$, with cost polynomial in $\frac{1}{\e}$ and $d$. This extends our earlier results on algorithms for estimating the ground state energy. The technique we present is sufficiently general to apply to
problems beyond the application studied in this paper.

The final publication is available at Springer via http://dx.doi.org/10.1007/s11128-015-0927-y.

\end{abstract}


\section{Introduction}
\label{sec:Intro}
Computing eigenvalues of Hamiltonians with a large number of degrees of freedom is a very challenging problem in computational science and engineering.  Hamiltonian eigenvalues give the system energy levels, corresponding to the \textit{ground} and \textit{excited} states. 
For example, one of the most important tasks in chemistry is to calculate the energy levels of molecules, which is required for predicting reaction rates and electronic structure, and which, in particular, depends principally on the low order energy levels.  The best classical algorithms known for such problems have cost that grows exponentially in the number of degrees of freedom \cite{Lanyon10}. Therefore, efficient quantum algorithms would be an extremely powerful tool for new science and technology.

On the other hand, there are a number of recent results in discrete complexity theory suggesting that many eigenvalue problems 
are very hard even for quantum computers because they are QMA-complete 
\cite{KempeLocalHam,intBosons,schuchDFT,childs2013bose}. However, discrete complexity theory deals with the worst case over large classes of Hamiltonians. It does not provide methods or necessary conditions determining when an  eigenvalue problem is hard. In fact, there is a dichotomy between theory and practice. As stated in  \cite{love2012back}, \lq\lq complexity theoretic proofs of the advantage of many widely used classical algorithms are few and far between.\rq\rq\
Therefore, it is important to develop new quantum algorithms and
to use them for solving  eigenvalue problems for which quantum computing can be shown to have a significant advantage over classical computing.

In \cite{GS} we developed an algorithm and proved strong exponential quantum speedup for approximating the ground state energy (i.e., the smallest eigenvaue) of the time-independent Schr\"odinger equation under certain assumptions. In \cite{Qspeedup} we explain why this problem is different from the QMA-complete problems of discrete complexity theory. In \cite{Convex} we relaxed an important assumption of \cite{GS} and extended our results to the ground state energy approximation for 
the time-independent Schr\"odinger equation with a convex potential. 

An important advance would be to obtain analogous results for approximating excited state energies under weakened assumptions. The techniques we have used previously for the ground state energy do not extend to excited state energies. Similarly, in computational chemistry, for instance, Hohenberg-Kohn density functional theory (DFT) is strictly limited to ground states \cite{Oliv05,HK1}. There are other flavors of DFT that may provide approximations of excited state energies. However, in general, approximate methods in computational chemistry often succeed in predicting chemical properties yet their level of accuracy varies with the nature of the species and may fail in important instances; see \cite{Aspuru-Guzik:2005yq, Lanyon10} and the references therein. Obtaining conditions allowing one to approximate excited state energies with a guaranteed accuracy and a reasonable
cost would provide a valuable insight into the compexity of these problems.

In this paper we present an entirely new approach for approximating a constant number of low order eigenvalues. At the same time we relax some of the assumptions of our previous work 
for approximating the ground state energy \cite{GS,Convex}. We will discuss these papers in Section \ref{sec:Overview}. Using the properties of our eigenvalue problem, we construct a set $\mathcal{S}$ of trial eigenvectors. This set contains vectors that overlap sufficiently with the unknown eigenvectors corresponding to the eigenvalues of interest. Then, these vectors can be used as initial states in quantum phase estimation (QPE) to produce eigenvalue estimates with a reasonably high (i.e., not exponentially small) success probability. The elements of $\mathcal{S}$ are known eigenvectors of a slightly perturbed problem. It is important to select the perturbation carefully so that the elements of $\mathcal{S}$ can be prepared efficiently on a quantum computer. In principle it is difficult to determine exactly which eigenvectors of the perturbed problem sufficiently overlap with the unknown eigenvectors of interest. Thus, our construction of $\mathcal{S}$ generally contains more elements than are absolutely necessary. Our algorithm runs QPE repeatedly with each element of $\mathcal{S}$ as initial state. We show that carefully selecting a constant number of the smallest measurement outcomes leads to estimates of the desired eigenvalues with a reasonable probability and cost, as long as the size of $\mathcal{S}$ is not exponentially large in the problem parameters. By reasonable cost we mean that the algorithm uses a number of qubits and quantum operations which is polynomial in the problem parameters. By reasonable success probability, we mean a probability $p$ that is bounded from below by a constant, e.g. $p \geq \frac34$. Unless the success probability of an algorithm is exponentially small in the problem parameters, it can be boosted to become arbitrarily close to $1$ using a number of repetitions that is also polynomial. We remark that the selection of the perturbation of the Hamiltonian impacts the size of $\mathcal{S}$, and hence the cost of our algorithm, and is an important consideration. 

We illustrate our results by considering the time-independent Schr\"odinger equation under weaker assumptions than those of \cite{GS,Convex}, as we explain below. For this problem, the cardinality of the set $\mathcal{S}$ of trial eigenvectors turns out to be polynomial in the number of degrees of freedom $d$. We derive cost and probability estimates for approximating a constant number 
of low order eigenvalues. Indeed, for accuracy $O(\e)$ the cost and the number of qubits of our algorithm is polynomial in $d$ and $\frac{1}{\e}$.
More precisely, we consider the eigenvalue problem
\begin{eqnarray}
\left(-\frac{1}{2}\Delta + V \right) \Psi(x) &=& E \: \Psi(x) \quad x \in I_d = (0,1)^d, 
\label{eqn:TISE1} \\ 
\Psi(x) &=& 0 \quad x \in \partial I_d,
\label{eqn:TISE2}
\end{eqnarray}
where $\Delta$ denotes the Laplacian and $\Psi$ is a normalized eigenfunction. Here all masses and the normalized Planck constant are set to one, and we assume the potential $V$ is a smooth and uniformly bounded function as we will explain later.

Our problem is to compute a constant number $j$ 
of estimates
$$\tilde{E}_0 < \tilde{E}_1 < ... < \tilde{E}_{j-1},$$ 
each approximating a different low order eigenvalue with error $O(\e)$ and high probability.
$\tilde{E}_0$ is the estimate of the ground state energy (i.e. the smallest eigenvalue). In general, the eigenvalues may be degenerate with unknown multiplicities. Moreover, eigenvalues may be clustered in balls of radius $O(\e)$. We call such eigenvalues $\e$-\textit{degenerate}. It is reasonable to assume that it is not necessary to produce estimates for every single $\e$-degenerate eigenvalue, and some of them can be omitted. 

Such eigenvalue problems can be solved by suitably discretizing the continuous operator (Hamiltonian) to obtain a symmetric matrix whose low order eigenvalues approximate those of the continuous problem, and then by approximating the matrix eigenvalues. Eigenvalue problems involving 
symmetric matrices are conceptually easy and methods such as the bisection method can be used to solve them with cost proportional to the matrix size, modulo polylog factors \cite{Demmel}. The
difficulty is that the discretization leads to a matrix of size that is exponential in $d$. Hence, the cost for approximating the matrix eigenvalue is prohibitive when $d$ is large. In fact, a stronger result is known, namely the cost of any deterministic classical algorithm approximating the ground state energy 
must be at least exponential in $d$, i.e., the problem suffers from the curse of dimensionality \cite{AP07}. It is important to point out that different approaches may lead to different matrix eigenvalue problems that have varying degrees of difficulty. For instance, in quantum chemistry, the first and second quantization approaches for computing energies of the electronic Hamiltonian, as described in \cite{Kassal}, lead to completely different matrices with different notions of degrees of freedom.  Moreover, discretizations of certain problems in physics may lead to eigenvalue problems for stoquastic matrices, that some believe are computationally easier to solve \cite{stoquastic}.

It is worth noting that in certain cases quantum algorithms may be able to break the curse of dimensionality by computing $\e$-accurate eigenvalue estimates with cost polynomial in $\e^{-1}$ and $d$. This was shown in \cite{GS, Convex} where we saw that for smooth nonnegative potentials that are uniformly bounded by a relatively small constant, or are  convex, 
there exists a quantum algorithm approximating the ground state energy with relative error $O(\e)$ and cost polynomial in $d$ and $\e^{-1}$.

It is important to investigate conditions for the potential $V$ beyond those of \cite{AP07,Convex,GS, Qspeedup}. In this paper we pursue this direction. 
As we indicated, we give a general algorithm for low order eigenvalues, and then apply it to the time-independent Schr\"odinger equation where $V$ is smooth and uniformly bounded by a constant. The algorithm has cost polynomial in $\e^{-1}$ and $d$, regardless of the size of the bound. 
We exhibit the resulting quantum algorithm, its cost, and success probability.
The technique that we have developed 
can be applied to other eigenvalue problems as well. 

We summarize the contents of this paper. 
In Section 2 we define our eigenvalue problem. 
We also review classical and quantum algorithms for eigenvalue problems. 
We discuss the limitations of classical algorithms, and how they may be overcome by quantum algorithms. 
We specify rather general conditions and provide a quantum algorithm which computes a constant number of approximations to low order eigenvalues, i.e. low order excited state energies (including the ground state). We explain how to construct a set $\mathcal{S}$ of trial eigenvectors for our algorithm using a perturbation approach. 
In Section 3, we study the overlaps between the trial eigenvectors and the unknown eigenvectors corresponding to the eigenvalues of interest. We provide lower bounds for the overlaps, and show how they depend on the cardinality of $\mathcal{S}$. In Section 4 we illustrate our results by considering a special case of the time-independent Schr\"odinger equation. This allows us to present specific estimates for the cost of our algorithm and its success probability, which we state explicitly in Theorem \ref{thm1}. 
Finally, we summarize our results in Section 5.

\section{Problem Definition}
\label{sec:ProbDef}
In this section, we introduce the eigenvalue problem in its most general form to emphasize that our approach applies under very broad conditions. In later sections, we will make more assumptions in order to show specific results. 

We consider an eigenvalue problem for a self-adjoint operator $L$ with 
a discrete spectrum. We will provide more details about $L$ below. Let 
\begin{equation}   \label{eq1}
E_{(0)} < E_{(1)} < ... < E_{(i)} < ...
\end{equation}
be its eigenvalues ignoring multiplicities, which we call the \textit{energy levels} of $L$. Suppose we want to estimate the lower part of the spectrum with accuracy $O(\e)$.
Since any two distinct eigenvalues of $L$ can be arbitrarily close to each other, any algorithm that approximates the lower part of the spectrum with accuracy $\e$ cannot be expected to distinguish between
all eigenvalues $E_{(k)} \neq E_{(l)}$ with $|E_{(k)} - E_{(l)} | = O(\e)$. We have called such eigenvalues $\e$-\textit{degenerate}. So the goal is to obtain an algorithm whose output will be $j$ numbers 
\begin{equation}   \label{eq:tildeE}
\tilde{E}_0 < \tilde{E}_1 < ... < \tilde{E}_{j-1}
\end{equation}
satisfying with high probability the following conditions:
\begin{enumerate}
\item[\bf{C1}] For every $i \neq k\in\{0,\dots,j-1\}$, there exist $E_{(s_i)}\neq E_{(s_k)}$ such that $|E_{(s_i)} - \tilde{E}_{i}|=O(\e)$ and $|E_{(s_k)} - \tilde{E}_{k}|=O(\e)$,  i.e. different outputs are approximations of different eigenvalues with error $O(\e)$, respectively.
\item[\bf{C2}]  If $|\tilde{E}_{i+1} - \tilde{E}_{i}|  = \omega(\e) $, there is no eigenvalue $E$ of $L$ satisfying $\tilde{E}_{i} < E < \tilde{E}_{i+1}$ and $\min(|\tilde{E}_{i+1} -E|, |\tilde{E}_{i} -E|) = \omega(\e)$.\footnote{For functions $f,g\ge 0$ defined on $\reals_+$, the notation $f(\e)=\omega(g(\e))$ means that for any $M>0$, arbitrarily large, we have $f(\e)\ge M g(\e)$ for sufficiently small $\e$.} Thus the algorithm doesn't miss (or skip) any eigenvalues in the lower part of the spectrum unless they are $O(\e)$ apart, i.e.,
{$\e$-degenerate}.
\end{enumerate}
Clearly, if $\e$ is sufficiently small such that the eigenvalues of $L$ are well-separated, then the algorithm produces approximations with error $O(\e)$ of the $j$ smallest distinct eigenvalues. 

Assume that $L^0$ is a self-adjoint operator defined on a separable Hilbert space,  $V$ is a symmetric operator whose domain contains the domain of $L^0$, and that $L = L^0 + V$ is self-adjoint on the domain, $D(L^0)$,  of $L^0$. We also assume that $L^0$ and $L$ have discrete spectra and that the eigenspaces associated with each eigenvalue are finite dimensional; see e.g. \cite{Gustafson,Sigal,Titchmarsh}.
In the general case, selecting the partition of $L$ to $L^0$ and $V$ is not trivial and may significantly affect the problem complexity; we do not deal with this problem here. Our discussion in this section applies equally well to Hermitian matrices.

Let 
\begin{equation}   \label{eqn:sigma}
\sigma \leq E_0 \leq E_1 \leq ... \leq E_i \leq ...
\end{equation}
be the eigenvalues of $L$ indexed in non-decreasing order, where $\sigma$ is a given lower bound. Ignoring possible eigenvalue multiplicities we have a strictly increasing subsequence of eigenvalues which we denote by 
\begin{equation} \label{eqn:exc}
E_{(0)} < E_{(1)} < ... < E_{(i)} < ...
\end{equation}
Similarly we denote by 
\begin{equation} \label{eqn:L0eval}
E_0^0 \leq E^0_1 \leq ... \leq E^0_i \leq ...
\end{equation} 
 the eigenvalues of $L^0$ indexed in non-decreasing order, and by 
\begin{equation} \label{eq:l0excited2}
E^0_{(0)} < E^0_{(1)} < ... < E^0_{(i)} < ...,
\end{equation}
the eigenvalues of $L^0$ ignoring multiplicities. 
Assume we know all the eigenvalues and eigenvectors of $L^0$. Often this is a reasonable assumption. For example, this is true for the eigenvalues and eigenvectors of the Laplacian $L^0=-\Delta$ defined on the $d$-dimensional unit cube with Dirichlet or Neumann boundary conditions.

We wish to estimate the low order eigenvalues of $L$. By \textit{low order} we mean that $j$ in equation (\ref{eq:tildeE}) is a constant. 
Intuitively, we expect a \lq\lq small\rq\rq\ and suitably well-behaved perturbation to have a proportionately \lq\lq small\rq\rq\  effect on the eigenvectors and eigenvalues of $L^0$. Algorithms solving this problem can take advantage of the known eigenvalues and eigenvectors of $L^0$.

\subsection{Background: Classical and Quantum Algorithms }     
\label{sec:Overview}
We briefly review algorithms for eigenvalue problems. 
Recall that we are interested in problems for which quantum computing can be shown to have a significant advantage over classical computing with performance guarantees in terms of accuracy and speed. Hence, we do not consider empirical approaches or heuristic eigenvalue algorithms. 

Algorithms approximating eigenvalues use a discretization of $L$ to obtain a matrix eigenvalue problem. For example, when $L$ is a differential operator, one can use a finite difference discretization \cite{Leveque}, or a finite element discretization \cite{strangFiniteElement, babuskaOsborn}. In particular, for the time-independent Schr\"odinger equation specified by equations  (\ref{eqn:TISE1}) and (\ref{eqn:TISE2}), a finite difference discretization has been used in \cite{GS, Convex}. Since $L$ is self-adjoint, the resulting matrix is symmetric.

Matrix eigenvalue problems have been extensively studied in numerical linear algebra, and there are classical algorithms for approximating one, or some, or even all of the eigenvalues and/or the corresponding eigenvectors of a matrix \cite{Demmel, Golub, Parlett, cullum2002lanczos}. Examples of such algorithms include the power method,  inverse iteration, the QR algorithm, and the bisection method for symmetric matrices. In particular, the bisection method for symmetric matrices can compute the eigenvalues that lie within a given range. In general, the known eigenvalues of $L^0$ can be helpful in computing such a range for the $j$ eigenvalues of $L$ that we consider in this paper. Typically, a symmetric matrix will be transformed to Hessenberg form, which is a 
tridiagonal matrix \cite[Sec. 4.4.7]{Demmel}. Then the bisection method is applied to the latter matrix \cite[Sec. 5.3.4]{Demmel}. 
The cost of this procedure, even if the original matrix is dense, is a 
low degree polynomial in the matrix size and $\log \frac{1}{\e}$, where $\e$ is the desired accuracy. Therefore, for matrices of moderate size, approximating the low order eigenvalues can be done at a reasonable cost. 
On the other hand, the costs of the above algorithms are bounded from below by a quantity that is at least proportional to the matrix size, even if the original matrix is sparse. Hence, the eigenvalue estimation problem becomes hard when the matrix size is huge.

Observe that the discretization of the operator $L$ must be sufficiently fine so that the eigenvalues of $L$ which are of interest are approximated by eigenvalues of the resulting matrix within the specified accuracy $\e$.  This increases the matrix size. For example, in the estimation of the ground state energy (smallest eigenvalue) of the time-independent Schr\"odinger equation, equations (\ref{eqn:TISE1}) and (\ref{eqn:TISE2}), with $V$ uniformly bounded by $1$, the finite difference discretization on a grid will yield a matrix of size 
$m^d \times m^d$, $m= 2^{\lceil  -\log_2 \e \rceil  } -1$ \cite{GS}. This means that the cost of the matrix eigenvalue algorithms mentioned above is bounded from below by a quantity proportional to  $\left(\frac{1}{\e}\right)^d$, i.e. the cost grows exponentially in $d$. In \cite{GS}, the Laplacian was discretized using a $2d+1$ stencil on a grid, and $V$ was discretized by evaluating it at the grid points. 

To approximate the ground state energy of the problem specified by equations (\ref{eqn:TISE1}) and (\ref{eqn:TISE2}) in the worst case with (relative) error $\e$, assuming $V$ and its first-order partial derivatives are uniformly bounded by $1$, and the function evaluations of $V$ are supplied by an oracle, a much stronger result holds.
The complexity (i.e., the minimum cost of any classical algorithm, and not just the eigenvalue algorithms mentioned above) is bounded from below by a quantity proportional to 
$\e^{-d}$ as $d\e \rightarrow 0$\cite{AP07, GS}. So unless $d$ is moderate, the problem is very hard and suffers from the curse of dimensionality. In \cite{Qspeedup}, we elaborate on this lower bound.
The same complexity lower bound applies to the approximation of low order eigenvalues under the same, or more general, conditions on $V$. Finally, we point out that the complexity of this problem in the classical randomized case is an open question. 

We now turn to quantum algorithms. There is a well-studied quantum algorithm, quantum phase estimation (QPE) \cite{AbramsLloyd, NC, Aspuru-Guzik:2005yq}, which can be used to approximate eigenvalues of a Hamiltonian $H$. More precisely, the algorithm approximates the phase corresponding to an eigenvalue of a unitary matrix, which in our case is $e^{-iH}$. QPE is efficient if two conditions are met. The first condition is that simulating a system evolving with Hamiltonian $H$ can be done efficiently, i.e. we can approximate $e^{-iHt}$, $t \in \mathbb{R}$, accurately with low cost. The second condition requires that 
we are given a relatively good approximation of an eigenvector corresponding to the eigenvalue of interest. In addition, one should be able to implement this approximation as a quantum state efficiently. The approximate eigenvector is used to form the initial state of QPE. We also remark that QPE uses the (quantum) Fourier transform as a module. The Fourier transform can be implemented efficiently on a quantum computer \cite{NC}.  

We discuss the two required conditions for QPE further. 
Simulating the evolution of a system under a Hamiltonian $H$ appears to be a difficult problem for classical computers when the size of $H$ is large. As proposed by Feynman \cite{Feynman}, quantum computers are able to carry out such simulation more efficiently in certain cases. For example, Lloyd \cite{Lloyd96} showed that local Hamiltonians can be simulated efficiently on a quantum computer. About the same time, Zalka \cite{Zalka1, Zalka2} showed that many-particle systems can be also be simulated efficiently on a quantum computer. Later, Aharonov and Ta-Shma \cite{AharonovTa-Shma} generalized Lloyd's results to sparse Hamiltonians. Berry et. al. \cite{Berry07} extended the cost estimates of \cite{AharonovTa-Shma}. The results of \cite{Berry07} were in turn improved by Papageorgiou and Zhang in \cite{PZ12}. Although there has been more work on quantum Hamiltonian simulation since then, the approach of \cite{Berry07, PZ12} suffices for our discussion. 
These papers assume that $H$ is given by a black-box (or oracle), and that $H$ can be decomposed efficiently by a quantum algorithm, using oracle calls, into a finite sum of Hamiltonians that individually can be simulated efficiently.
In this paper, where $L=L^0 + V$, we assume that the Hamiltonians resulting from the discretizations of $L^0$ and $V$ can be simulated efficiently on a quantum computer. Their sum, i.e. the Hamiltonian obtained from the discretization of $L$, can be simulated efficiently using splitting formulas such as the Trotter formula, the Strang splitting formula, or Suzuki's high-order splitting formulas. Simulation cost estimates are shown in \cite{PZ12}. 

Moreover, since we know the eigenvalues and eigenvectors of $L^0$, in certain cases one might be able to simulate its evolution explicitly without relying on an oracle. For instance when $L^0 = -\Delta$, a quantum algorithm and circuit implementing the evolution of the discretized Laplacian, is shown in \cite{Poisson}. That paper deals with the solution of the Poisson equation with Dirichlet boundary conditions. The efficient simulation of the discretized Laplacian is achieved by diagonalizing it using the quantum Fourier transform. The efficient simulation of the discretization of $V$ (which is a diagonal matrix) is achieved using an oracle and quantum parallelism. 

We now turn to the second requirement of QPE, namely the availability of a good approximate eigenvector. 
QPE will produce an estimate of the eigenvalue $\lambda$ (or more precisely, an estimate of the phase $\phi \in [0,1)$ corresponding to $\lambda$ through $\lambda = e^{2\pi i\phi}$) with success probability proportional to the quality of the approximate eigenvector \cite{NC, AbramsLloyd}. If the eigenvector providing the initial state of QPE 
is known exactly, the parameters of QPE can be set so its success probability is arbitrarily close to 1 \cite{NC}. If, on the other hand, we use an approximate eigenvector, the success probability is reduced proportionally to the square of the magnitude of the projection of the approximate eigenvector onto the actual eigenvector (i.e. the square of the \textit{overlap} between the two vectors)  \cite{AbramsLloyd}.  As long as this overlap is not exponentially small, QPE is efficient. 

For example, in \cite{GS, Convex} we show quantum algorithms that meet the two requirements of QPE and 
approximate the ground state energy for special cases of the time-independent Schr\"odinger equation, as specified in equations (\ref{eqn:TISE1}) and (\ref{eqn:TISE2}). In \cite{GS}, we assumed $V$ and its first-order partial derivatives are uniformly bounded by $1$. In that paper we show a quantum algorithm estimating the ground state energy with cost polynomial in $\frac{1}{\e}$ and $d$. In \cite{Qspeedup}, we explain why quantum algorithms have a significant advantage over classical algorithms solving this problem in the worst case. In the second paper \cite{Convex}, we extend the results to a different class of potential functions $V$, namely convex functions uniformly bounded by an arbitrary constant $C>1$ with first-order partial derivatives uniformly bounded by a constant $C'$. Under these assumptions, we derive a multistage quantum algorithm for estimating the ground state energy. The convexity of $V$, along with a recent result \cite{andrews2011proof} concerning the fundamental gap  (i.e., the difference between the first two eigenvalues) of Schr\"odinger operators, allows us to use a number of stages that is polynomial in $d$, with each stage having cost polynomial in $\frac{1}{\e}$ and $d$, so that the overall algorithm is efficient. Therefore, 
for special cases of the time-independent Schr\"odinger equation,
quantum algorithms vanquish the curse of dimensionality. 

We remark that obtaining a good approximate eigenvector required for QPE is a particularly difficult task, in general, when the matrix size is huge. Things are complicated further if one needs a number of different approximate eigenvectors, in order to use QPE to approximate the $j>1$ lower order eigenvalues. We overcome this difficulty for $L=L^0 + V$ using the known eigenvalues and eigenvectors of $L^0$, and properties of $V$, as we discuss below.

\subsection{Algorithm}
\label{sec:Algorithm}

\subsubsection{Algorithm: Idea}
\label{sec:AlgorithmIdea}

Our goal is to use QPE to estimate $j$ low order energy levels of $L$. For this, we need relatively good approximations of the corresponding eigenvectors. Since $L=L^0+V$ (i.e. $L$ and $L^0$ differ by the perturbation $V$), and since we know the eigenvalues and the eigenvectors of $L^0$, we can use them to obtain the necessary approximate eigenvectors. We indicate how this can be done. For simplicity and notational convenience, we do not distinguish between operators and their matrix discretizations in the rest of this section, since it is not important for the moment. Let $E$ denote one of the low order eigenvalues of $L$ (see, equation (\ref{eqn:exc})) that we wish to estimate. Intuitively, we expect a \lq\lq small\rq\rq\ and suitably well-behaved perturbation to have a proportionately \lq\lq small\rq\rq\  effect on the eigenvectors and eigenvalues of $L^0$. Let $u$ be an arbitrary unit vector belonging to the 
eigenspace associated with $E$. Then  there exists an eigenvector $u_k^0$ of $L^0$, similarly corresponding to a low order eigenvalue, that has an \textit{overlap} (magnitude of projection) with $u$ that is non-trivial, i.e.  $| \braket{u_k^0}{u}|$ is not extremely small, as we will see later.

A simple illustration of this idea is to imagine an $L^0$ with a symmetric ground state, and a perturbation $V$ that is relatively asymmetric. In such a case, the ground state of $L$ may overlap primarily not with the ground state eigenvector of $L^0$ but with an excited state. For example, the ground state of the one-dimensional harmonic oscillator is a (spatially) even function about the center of the well, and successive eigenstates are alternately odd and even. An odd perturbation of sufficient size will cause the ground state to become approximately odd, and project increasingly onto an odd unperturbed eigenstate.

An important idea in this paper is to form a collection $\mathcal{S}$ of the eigenvectors of $L^0$ that correspond to eigenvalues of $L^0$ that satisfy a certain property, which we specify in the next subsection. 
The goal is to have at least one element in $\mathcal{S}$ that has a reasonable overlap with a vector in the eigenspace corresponding to $E_{(i)}$, for each $i=0,1,...,j-1$. We call $\mathcal{S}$ the \textit{set of trial eigenvectors}. We will use each one of the elements of $\mathcal{S}$ repetitively as initial state in QPE, running QPE multiple times, to obtain a sequence of approximations that will lead us to estimates of each $E_{(i)}$.

Let us briefly discuss the idea for constructing $\mathcal{S}$. At one extreme, one could take $\mathcal{S}$ to be all of the eigenvectors of $L^0$, because not all of them have a negligible overlap with the eigenvectors of $L$ corresponding to the eigenvalues of interest. However, then the size of $\mathcal{S}$ can be huge. To limit $|\mathcal{S}|$, we select eigenvectors of $L^0$  that correspond to eigenvalues that do not exceed a certain bound. 
Roughly speaking, we will be excluding eigenvalues of $L^0$ that correspond to energies grossly exceeding the energies of $L$ that we wish to estimate. This idea is made precise in equation ($\ref{eq:upperBound}$) in next section.

The cardinality of $\mathcal{S}$ depends on the eigenvalue distribution of $L^0$.  
If the cardinality of $\mathcal{S}$ is not prohibitively large, and if we can discretize its elements and  efficiently prepare the corresponding quantum states, then we can run QPE repeatedly for the all elements of $\mathcal{S}$ to produce an estimate of $E_{(i)}$ among its different outputs with a sufficiently high probability, $i=0,1,...,j-1$. This probability can be boosted to become arbitrarily close to $1$ using further repetitions of the procedure. We remark that the cardinality of $\mathcal{S}$ depends on the distribution of eigenvalues of $L^0$ and the properties of $V$.
Observe that detecting the desired estimates $\tilde{E}_0, ..., \tilde{E}_{j-1}$ from the outcomes obtained from the different runs of QPE is not a trivial task, and we will show how this is accomplished.

\subsubsection{Algorithm: Description}
\label{sec:AlgorithmDescrip}
Let $V$ be such that $\| V\|_{L^0}:= \sup\{ \|Vu\|: u \in D(L^0), \|u\| = 1\} < \infty$ uniformly in $d$. 
 Assume we are given (or we have derived) $c>1$ 
and $B$ a sufficiently large upper bound on the lower part of the spectrum of $L$ which is of interest.\footnote{We give an explicit construction for $B$ in equation (\ref{hyp1}).} 
Consider the set of indices 
\begin{equation} \label{eq:upperBound}
\mathcal{I}:=\{i:E^0_i-B > c\|V\|_{L^0} \} \neq \emptyset.
\end{equation}

We define $\mathcal{S}$ to be the set of eigenvectors of $L^0$ that correspond to eigenvalues $E_i^0$ with $i \notin \mathcal{I}$;
in the case of degeneracy, it suffices to select any basis of the degenerate subspace. By constructing $\mathcal{S}$ in this way, we are guaranteed that at least one of its elements will  overlap sufficiently with an element of the degenerate eigenspace corresponding to each $E_{(i)}$, for $i=0,1,...,j-1$.
We will show that the magnitude of this overlap is bounded from below by a positive constant. 

The following is an overview of our quantum algorithm for approximating $j=O(1)$ low order eigenvalues of the operator $L=L^0+V$. Algorithm 1 deals with the special case of approximating the ground state energy $E_{(0)}$.
This algorithm illustrates our idea of using a set of trial eigenvectors to approximate an eigenvalue of $L$. 
Algorithm 2 computes the sequence of approximations $\tilde{E}_{1}, \tilde{E}_{2},...,\tilde{E}_{j-1}$, where each $\tilde{E}_{i}$ is computed using the values $\tilde{E}_{0}$ through $\tilde{E}_{i-1}$. Thus the overall procedure consists of iterating Algorithm 2 
until we obtain the $j$ desired estimates of equation $(\ref{eq:tildeE})$.

Let us pretend for the moment that $L^0$ and $L$ are $N\times N$ matrices; we do this for notational convenience. In later sections, when we consider specific instances of $L^0$ and $L$, we will show how to discretize them and obtain symmetric matrices such that each of the low order eigenvalues of these matrices approximates the corresponding eigenvalue of the respective operator with error proportional to $\e$. 

Our algorithms are based on QPE and require two quantum registers. 
The first register (\textit{top} register) contains sufficiently many qubits $t$ to guarantee the required accuracy $O(\e)$ in the results with a reasonable success probability for QPE. 
The second register (\textit{bottom} register) contains 
the necessary number of qubits to hold an approximate eigenstate.
\vskip 1pc 
\textbf{Algorithm 1. Description - Ground State Energy:}
\begin{enumerate}
     \item Define $\mathcal{S}$, the \textit{set of trial eigenvectors}, to be all eigenvectors of $L^0$ that correspond to eigenvalues $E^0_i$ $i\notin \mathcal{I}$ as defined in equation (\ref{eq:upperBound}). We denote these eigenvectors by  $u_k^0$ for $k=0,1,.., |\mathcal{S}|-1$.
    \item Set k = 0.
    \item Prepare the initial quantum state $\ket{0}^{\otimes t} \ket{u_k^0}$. The value of $t$ is chosen so that QPE, with relatively high probability, produces outcomes leading to energy estimates with error $O(\e)$. 
   \item Perform QPE with initial state $\ket{0}^{\otimes t} \ket{u_k^0}$ using the unitary matrix $U = e^{ i A/ R}$. $A =  L$ if $L$ non-negative definite, and otherwise $A= L - \sigma I $, where $\sigma$ is a lower bound to the minimum eigenvalue of $L$ as we have assumed in the previous section. The parameter $\sigma$ is assumed to be known; see equation (\ref{eqn:sigma}). Nevertheless, even if $\sigma$ is not known, it is often possible to obtain a convenient estimate of $\sigma$ using the eigenvalues of $L^0$ and the properties of $V$.   The goal is that $A$ is a non-negative definite matrix. The parameter $R$ is an upper bound to the spectral norm of $A$, which can be obtained using the eigenvalues of $L^0$ and the properties of $V$. The purpose of $R$ is to ensure that the resulting phases will lie in the interval $[0,1)$.
   \item Measure the first $t$ qubits, which give the result of QPE, and store the resulting value classically. We assume that the measurement outcomes are truncated to $b$ bits and we obtain non-negative integers in the range $\{0,...,2^b-1 \}$, where $b<t$. The role of the extra qubits $t_0=b-t$ is to increase the success probability of QPE.
    \item $k\leftarrow k+1$.
   \item Repeat steps 3-6 while $k < |\mathcal{S}|$. 
    \item Repeat steps 2-7 $r$ many times, where $r$ is a number precomputed to ensure with high probability that the stored results after $r$ runs contain an estimate of $E_{0}$. The value of $r$ depends on the problem at hand. In Section \ref{sec:TISE} we derive $r$ for a particular application.
    \item Take the minimum value of the stored measurement outcomes, mark it as selected, and convert it to an eigenvalue estimate $\tilde{E}_{0}$ of $E_{0}$ using the values of $\sigma$ and $R$ in the definition of $A$.
    \item Output $\tilde{E}_{0}$.
\end{enumerate}
\vskip 1pc
Note, the purpose of step 7 is to run QPE $|\mathcal{S}|$ many times, once with each $\ket{u_k^0}\in \mathcal{S}$ as input, because we do not know which of the elements of $\mathcal{S}$ has the largest overlap with the unknown ground state eigenvector, and the success probability of each run depends on this overlap. Since the largest overlap between the elements of $\mathcal{S}$  and the unknown eigenvector may not be sufficiently large so that the resulting success probability of the algorithm is bounded from below be a constant, say $\frac34$, the purpose of step 8 is to repeat the entire procedure $r$ many times to boost the success probability of computing $\tilde{E}_{0}$ correctly.

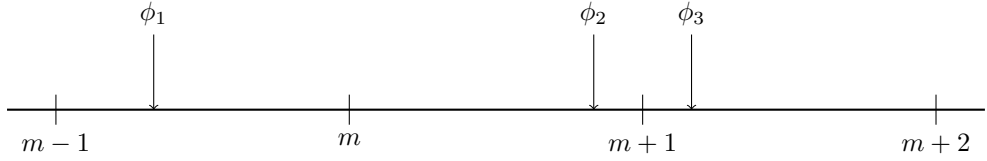
\begin{figure*}
\centering
\begin{tikzpicture}[xscale=1.3]
\draw [->](1,1) -- (1,0);
\draw [->](5.5,1) -- (5.5,0);
\draw [->](6.5,1) -- (6.5,0);
\node [above] at (1,1) {$\phi_1$};
\node [above] at (5.5,1) {$\phi_2$};
\node [above] at (6.5,1) {$\phi_3$};
\draw [thick] (-0.5,0) -- (9.5,0);
\draw (0,-.2) -- (0, .2);
\draw (3,-.2) -- (3, .2);
\draw (6,-.2) -- (6, .2);
\draw (9,-.2) -- (9, .2);
\node [below] at (0,-0.2) {$m-1$};
\node [below] at (3,-0.2) {$m$};
\node [below] at (6,-0.2) {$m+1$};
\node [below] at (9,-0.2) {$m+2$};
\end{tikzpicture}
\caption{Example of the selection of the measurement outcomes. Consider three phases $\phi_1$, $\phi_2$ and $\phi_3$ as shown. Assume that the distance between possible outcomes corresponds to error $O(\e)$. 
If $m-1$ is selected to estimate $\phi_1$, the next possible outcome the algorithm selects is $m+1$, which provides an estimate for both $\phi_2$ and $\phi_3$ in this example. Alternatively, if $m$ is selected to provide an estimate of  $\phi_1$, the next possible outcome is $m+2$, which provides an estimate of $\phi_3$, and the algorithm does not care to produce a separate estimate for $\phi_2$. Note that $\phi_2$ and $\phi_3$ are $\e$-degenerate and either can be ignored.}
\label{Fig:phaseEstimation}      
\end{figure*}

The following iterative algorithm extends Algorithm 1 to compute the sequence of approximations $\tilde{E}_{1}, .. , \tilde{E}_{j-1}$  satisfying the conditions of equation (\ref{eq:tildeE}), respectively. Every term of the computed sequence depends on all of the previously computed terms.
\vskip 1pc 
\textbf{Algorithm 2. Description - Excited State Energies:}
\begin{enumerate}
  \item Consider $A$ as defined in Algorithm 1. Run Algorithm 1 and let $\tilde{E}_{0}$ be its output.
  \item Set $i=1$ and prepare to compute an estimate of $\tilde{E}_{1}$.
  \item Repeat steps 1-8 of Algorithm 1, storing the outcome of every measurement. We assume that the measurement outcomes are truncated to $b$ bits and we obtain non-negative integers in the range $\{0,...,2^b-1 \}$, where $b<t$. The role of the extra qubits $t_0=b-t$ is to increase the success probability of QPE.
  \item Take the minimum of the measurement outcomes that exceeds by $2$  the last selected outcome and mark it selected. This way, with high probability, for each eigenvalue the error will be $O(\e)$, and the algorithm will not produce two different estimates for the same eigenvalue. 
Note that by  taking the minimum outcome relative to the previously selected outcome implies that the algorithm does not fail to produce estimates for consecutive eigenvalues, unless the eigenvalues differ by $O(\e)$. 
See Figure~\ref{Fig:phaseEstimation}.
  \item Use the values of $\sigma$ and $R$ in the definition of $A$ to rescale and shift the newly selected outcome to obtain the estimate $\tilde{E}_i$.
  \item Set $i\leftarrow i+1$ and prepare to compute the estimate $\tilde{E}_i$. 
  \item Repeat steps 3 through 6 if $i < j$.  
  \item Output $\tilde{E}_{1}, ..., \tilde{E}_{j-1} $.
\end{enumerate}
\vskip 1pc


It is clear that our procedure as outlined by Algorithms 1 and 2 will produce
the $j$ desired estimates (\ref{eq:tildeE}) of equation (\ref{eq1}). However, its cost varies depending on $V$ and the distribution of eigenvalues of $L^0$, which as we already mentioned determine the cardinality of $ \mathcal{S}$.  In the next sections we derive tight estimates for the cost and the success probability of our algorithm for particular choices of $L^0$ and $L$.

We remark that in cases where $L$ and $L^0$ are given explicitly, using their properties one may be able to obtain a set of trial eigenvectors $\mathcal{S}$ with significantly smaller cardinality, 
substantially improving the cost of the algorithm. For example, knowledge of the symmetry groups of $L$, $L^0$, and $V$ could be used to immediately rule out candidate eigenvectors. It is important to observe that different partitionings of the Hamiltonian into $L^0$ and $L-L^0$ may lead to very different sets of trial eigenvectors. Given a Hamiltonian $L$, an important task is to select $L^0$ that will result in a relatively small set of trial eigenvectors which can be computed efficiently.

\section{Preliminary Analysis} 
\label{sec:PrelimAnal}

Consider operators $L^0$ and $L$ with the assumptions of Section \ref{sec:Overview}. Recall that  $L^0$ and $L$ are self-adjoint operators on a Hilbert Space $\mathcal{H}$ with discrete spectra; e.g., see \cite{Gustafson}. 
Then we have the following eigenvalue equations:
$$L^0 u_i^0 = E^0_i u_i^0 ,\quad i = 0,1,2...  ,{\rm\ and\ let\ } E^0_0\leq E^0_1\leq E^0_2 \leq ...$$
$$L u_i = E_i u_i ,\quad i = 0,1,2...  ,{\rm\ and\ let\ } E_0\leq E_1\leq E_2 \leq ...$$
Without loss of generality we take all eigenvectors to have unit length. We assume that the eigenpairs $\{E_i^0, u_i^0\}_{i=0}^\infty$ are known. 

First suppose we wish to compute a specific eigenvalue $E$ of $L$. Let $u$ be a unit vector in the (possibly degenerate) subspace associated with $E$. We have
$$ \| L u - L^0 u\|^2  =  \| (L^0 + V) u - L^0 u\|^2 = \|V u \|^2 
\leq  \|V\|_{L^0}^2 ,$$
since $u$ belongs to the intersection of the domains of $L^0$ and $L$.
Expanding in the basis of unperturbed eigenvectors, we have $\ket{u} = \sum_i \beta_{i} \ket{u^0_i}$ where $\beta_{i} = \braket{u^0_i}{u}$. Then
$$ \| L u - L^0 u\|^2  = \| E \sum_i \beta_{i} \ket{u^0_i} - \sum_i \beta_{i} E_i^0 \ket{u^0_i}  \|^2 = \| \sum_i \beta_{i} (E - E_i^0)  \ket{u^0_i} \|^2$$
Combining these expressions and using the eigenvector orthonormality gives
\begin{equation} \label{eqn:sum}
 \|V\|_{L^0}^2 \geq \| L u - L^0 u\|^2  = \sum_i |\beta_{i}|^2 (E^0_i - E)^2 
\end{equation}
Assume that there exists $c>1$ such that the condition of equation (\ref{eq:upperBound}) holds. Observe that this is true for instances of the time-independent Schr\"odinger equation \cite{Titchmarsh}. 
From equation (\ref{eqn:sum}), using $B \geq E$, we obtain
$$ \|V\|_{L^0}^2 \geq  \sum_{i \in \mathcal{I} } |\beta_{i}|^2 (E_i^0 - E)^2 \geq  \sum_{i \in \mathcal{I}} |\beta_{i}|^2 (c\|V\|_{L^0} )^2 $$
which we rearrange as
\begin{equation}
  \sum_{i \in \mathcal{I}} |\beta_{i}|^2   \leq \frac{1}{c^2}  
\end{equation} 
or equivalently 
\begin{equation} \label{eqn:probBound}
 \sum_{i \notin \mathcal{I}} |\beta_{i}|^2  \geq 1 - \frac{1}{c^2  }  =: q   
\end{equation}
Thus there must exist an index $k \notin \mathcal{I}$ such that $|\beta_k|^2 \geq \frac{q}{|\mathcal{S}|}$. If $|\mathcal{S}|$ is not extremely large, then one of the first 
$|\mathcal{S}|$ eigenvectors of $L^0$ must have a reasonable overlap\footnote{Here and elsewhere, by reasonable overlap we mean that the magnitude of the projection is not exponentially small in $d$.}  with $u$.

\section{Application: Time-Independent Schr\"odinger Equation} 
\label{sec:TISE}
In this section we consider the time-independent Schr\"odinger equation on the $d$-dimensional unit cube with Dirichlet boundary conditions to illustrate the algorithms of Section \ref{sec:Algorithm} that compute the $j$ estimates of equation (\ref{eq:tildeE}), where $j=O(1)$. In particular, consider the eigenvalue problem
\begin{eqnarray}   \label{eqn:eqn}
L u (x) := (-\tfrac 12 \Delta +V)u(x) &=& Eu(x)\quad \mbox{for all}\;\; x\in I_d:=(0,1)^d, \\
u (x) &=& 0 \quad \mbox{for all}\;\; x\in  \partial I_d, \nonumber
\end{eqnarray}
where  $V$ is uniformly bounded by a constant $M$ and has continuous first-order partial derivatives in each direction uniformly bounded by a constant $C$, i.e. $\| \frac{\partial}{\partial x_i} V\| \leq C$. 
Thus, without loss of generality we assume that $V\geq 0$.
We set $L^0=-\frac{1}{2} \Delta$, where
$$ \Delta = \sum_{i=1}^d \frac{\partial^2}{\partial x_i^2} .$$
We assume that the eigenvalues of $L$ and $L^0$ are indexed in non-decreasing order. We want to approximate the first $j$ excited state energies, $E_{(0)},\dots, E_{(j-1)}$ , (i.e. the $j$ smallest eigenvalues ignoring multiplicities) with error proportional to $\e$, modulo $\e$-degenerate eigenvalues as explained previously. 
Thus we are interested in low order excited state energies because we have assumed that  $j$ is a constant.
Recall our definitions and notation of Section \ref{sec:ProbDef}. 

As we already indicated, the cost of Algorithms 1 and 2 depends on the cardinality of a set of trial eigenvectors $\mathcal{S}$. 
We will now show that $|\mathcal{S}|$ is bounded by a polynomial in $d$
 in the case  of the Schr\"odinger equation we are considering here.

The eigenvalues and eigenvectors of $L^0 = -\frac{1}{2}\Delta$ are known to be
\begin{equation} \label{eqn:eval}
 E_{\vec{k}}^0 =  \frac12 (k_1^2 + k_2^2 + \hdots+ k_d^2)\pi^2 \;\;\;\;  \; \vec{k}=(k_1,...,k_d) \in \mathbb{N}^d
\end{equation}
$$ u_{\vec{k}}^0 (\vec{x})= 2^{d/2} \prod_{i=1}^d \sin(k_i \pi x_i) \;\;\;\;  \vec{x}=(x_1,...,x_d) \in [0,1]^d \;\;\;\; \vec{k}=(k_1,...,k_d) \in \mathbb{N}^d.$$ 
We may re-index them by considering the eigenvalues  in non-decreasing order to obtain $E^0_0\leq E^0_1 \leq ... \leq E_i^0 \leq ...$ as in (\ref{eqn:L0eval}). Thus $E^0_0 = \frac12 d\pi^2 < \frac12 (d+3)\pi^2 = E^0_1 $ and $E^0_1 $ is a degenerate eigenvalue with dimension of its associated degenerate subspace equal to $d$. Similar considerations apply to the rest of the eigenvalues. We remark that the distribution of the eigenvalues of $L^0$ is known \cite{Titchmarsh}.

We will use (\ref{eq:upperBound}) with $c=2$ to derive a set of trial eigenvectors and bound its cardinality. 
In fact, we derive a set of trial eigenvectors that is slightly larger than the set obtained by strictly considering the indices in the complement of $\mathcal{I}$ in (\ref{eq:upperBound}). Yet its size is polynomial in $d$ as we will see, and for the sake of brevity, we also denote this set by $\mathcal{S}$. In particular, we construct the quantity $B$ of equation (\ref{eq:upperBound}) and show a $K=K(j,V)$ such that for $k \geq K  \Rightarrow E_k^0 > 2M + B$. So we obtain an upper bound for the $jth$ largest eigenvalue of $L$.
Clearly the cardinality of $\mathcal{S}$ grows with $B$ because we include eigenvectors of $L^0$ that correspond to increasingly large eigenvalues. The purpose of the construction below is to obtain a crude but helpful in our analysis estimate of the distribution of the eigenvalues of $L$ using the eigenvalues of $L^0$; in particular to cover possible degeneracy of the eigenvalues of $L$.

We select $j+1$ values $E^0_{(s_n)}$ from the strictly increasing sequence of eigenvalues (see  (\ref{eq:l0excited2})) such that 
\begin{equation}  \label{eq:subseq}
E^0_{(s_0)}:= E^0_{(0)}  < E^0_{(0)} + M < E^0_{(s_1)} < E^0_{(s_1)} +M <  \hdots  < E^0_{(s_{j-1})} +M < E^0_{(s_j)} 
\end{equation}
where $E^0_{(s_n)} - E^0_{(s_{n - 1})} > M$, $n=1,\dots,j$. Indeed it is possible to select a subsequence that satisfies these conditions. 
We know that $E^0_{(s_{n-1})} = \frac{1}{2}(k_1^2 + k_2^2 +... k_d^2)\pi^2$ for a certain $\vec{k}$. The inequality 
$$\frac12 ({k'}_1^2 + {k'}_2^2 +... {k'}_m^2)\pi^2 + \frac12 ({k}_{m+1}^2 + {k}_{m+2}^2 +... {k}_d^2)\pi^2  \geq E^0_{(s_{n-1})} + M$$
is satisfied by selecting $m$ to be a suitable constant and then by selecting $k'_i \geq k_i + \gamma_i$, where $\gamma_i$ is a suitable positive integer constant, $i=1,\dots, m$.
For example, after fixing $m$, we can repeatedly increment each of the $k'_i$, $i\in\{1,\dots,m\}$, successively until the desired inequality holds. 
Iteratively, we define $E^0_{(s_n)} = ({k'}_1^2 + {k'}_2^2 +... {k'}_m^2)\pi^2/2 + ({k}_{m+1}^2 + {k}_{m+2}^2 +... {k}_d^2)\pi^2/2$ for $n=1,2,..,j$.

By our construction, the interval $[E^0_{(s_0)},E^0_{(s_j)}]$ contains at least $j$ distinct eigenvalues of $L$, %
since $E^0_{(s_j)} - E^0_{(s_0)} > jM$, and for every $i$, $E_i^0 \leq E_i \leq E_i^0 + M$. Moreover, $E^0_{(s_j)}  = E_0^0 + c'$, where $c'$ is a constant. Thus, we 
take $c=2$ in (\ref{eq:upperBound}) and
define the constant $B$ as
\begin{equation} \label{Bdef}
B:= M + E_{(s_j)}^0 = M + E_{0}^0 + c' .
\end{equation}
From (\ref{eqn:eval}) there exists a $K\in\nat$ such that
\begin{equation} \label{hyp1}
k\ge  K\;\; \Rightarrow \;\;E_k^0 > 2M +B  = 3M + E_{(s_j)}^0 .
\end{equation}
Hence, we construct the \textit{set of trial eigenvectors} $\mathcal{S}$ to be the set of all eigenvectors of $L^0$ that correspond to eigenvalues less than or equal to $3M + E_{(s_j)}^0$. We bound $|\mathcal{S}|$ next. 

The cardinality of  $\mathcal{S}$ is the number of tuples $\vec{k}\in \nat^d$ such that $(k_1^2 + \dots + k_d^2)\pi^2/2 \le 3M + d\pi^2/2 + c' $.
Let $m$ be the number of components $k_{i_1}, ..., k_{i_m}$ of such a $\vec{k}$ that are greater than $1$. Then we have 
$$(d-m)\pi^2/2 +(k_{i_1}^2 +... k_{i_m}^2)\pi^2/2 \le 3M + d\pi^2/2 + c' .$$
Since $k_i \geq 2$ we have
$$3m\pi^2 \leq -m\pi^2 +(k_{i_1}^2 +... k_{i_m}^2)\pi^2 \le  2(3M + c') .$$
Hence, $m$ is $O(1)$.  Therefore, in order to construct $\mathcal{S}$ one needs to consider tuples $\vec{k}'\in\nat^d$ where at most a constant number of components are greater than $1$.
The number of such tuples depends on the number of possible combinations by which  one can select a constant number of components of $\vec{k}'$ to be greater than or equal to~$2$. Therefore, this number is polynomial in $d$.\footnote{This follows immediately for $m=O(1)$ from the bound $\binom{d}{m}\leq \frac{d^m}{m!} = poly(d)$.}

Table \ref{tab:tab1} below shows the eigenvalues of the Laplacian by considering tuples where a constant number $m$ of components exceed $1$, assuming that these components are each bounded by a constant $N$.
Observe that in all cases, since $m=O(1)$, the multiplicity of the eigenvalues is polynomial in $d$.


\begin{table}
\centering 
\caption{Distribution of eigenvalues of $L^0 = -\frac12 \Delta$ with respect to the number $m$ of indicies $k_i \geq 2$. }
\label{tab:tab1}   
\begin{tabular}{c c c}  
\hline\noalign{\smallskip} 
$m$ & Combinations & Eigenvalue \\ [0.5ex] 
\hline\noalign{\smallskip} 
$0$ & $\binom{d}{0}$ & $d\pi^2/2$ \\
$1$ & $\binom{d}{1}$ & $(d-1)\pi^2/2 + k_{i_1}^2\pi^2/2$ \\
$2$ & $\binom{d}{2}$ &  $ (d-2)\pi^2/2 + (k_{i_1}^2+k_{i_2}^2)\pi^2/2$  \\
$...$\\
$l\leq d$ & $\binom{d}{l}$ & $(d-l)\pi^2/2 + (k_{i_1}^2+..+k_{i_l}^2)\pi^2/2$ \\
[1ex] 
\hline\noalign{\smallskip} 
\end{tabular}
\end{table}

Therefore, the cardinality of $\mathcal{S}$ is polynomial in $d$. As shown in Section \ref{sec:PrelimAnal}, for every eigenvector of $L$ that corresponds to an eigenvalue less than or equal to $B$, there exists an eigenvector of $L^0$ in  $\mathcal{S}$ such that the two eigenvectors have a non-trivial overlap and it follows from (\ref{eqn:probBound}) that the magnitude squared of this projection of the one onto the other will be at least $\frac{q}{poly(d)} = \frac34 \frac1{poly(d)} = O(\frac{1}{poly(d)} )$.

\subsection{Finite Difference Discretization}  
\label{sec:FiniteDiff}
We obtain a matrix eigenvalue problem by discretizing  (\ref{eqn:eqn}) on a grid with mesh size $h= \frac{1}{N+1}$, $N\in \nat$, using finite differences \cite{Leveque, GS}.
This yields a matrix $M_h:= -\tfrac{1}{2}\Delta_h + V_h$ with size $N^d \times N^d$. The matrix $-\frac12\Delta_h$ is obtained using a $2d+1$ stencil for the Laplacian \cite[p.60]{Leveque}. 
It is known that the low order eigenvalues of $M_h$ approximate the corresponding eigenvalues of $L$. The eigenvalues and eigenvectors of $-\tfrac{1}{2} \Delta_h$ are known and are given by
\begin{equation}
E^0_{h,\vec{k}} = \frac{2}{h^2} \sum_{i=1}^d \sin^2(\pi h k_i /2) \;\;\;\;  \;\vec{k}=(k_1,...,k_d) \;\;\;\; 1\leq k_i \leq N
\end{equation}
\begin{equation}
u^0_{h,\vec{k}} = \bigotimes_{i=1}^d v_{k_i},
\end{equation} 
where the vectors $v_{k_i}\in \mathbb{R}^d$ have coordinates
\begin{equation} \label{eq:discreteEvecs}
v_{k_i, \ell} = \sqrt{2h} \sin(k_i \ell \pi h ) \;\;\;\;  \;\ell = 1,2,...,N \;\;\;\;  \;i= 1,2,...,d.
\end{equation}
Similarly to (\ref{eq:l0excited2}), we index the eigenvalues of $-\tfrac{1}{2} \Delta_h$ in increasing order ignoring multiplicities to obtain
\begin{equation} \label{eq:l0excited}
E^0_{h,(0)} < E^0_{h,(1)} < ... < E^0_{h,(i)} < ...
\end{equation}
Then from \cite{Weinberger1} we have
\begin{equation}  \label{eq:cdh2}
|E^0_{h,(k)}-E^0_{(k)}| \leq Cdh^2    \;\;\;\;  \;\ \text{ for } k = O(1)
\end{equation}
where $C>0$ is a constant.

$V_h$ is an $N^d \times N^d$ diagonal matrix which contains evaluations of $V$ at the grid points truncated to $\lceil  \text{log}_2 h^{-1} \rceil $ bits of accuracy. Thus $M_h$ is symmetric, positive definite, and sparse. This matrix has been extensively studied in the literature \cite{Demmel, FW, Leveque}. 
For $V$ that has bounded first-order partial derivatives and $k=O(1)$, using the results of \cite{Weinberger1, Weinberger2} we have that 
there exists a matrix eigenvalue $E_{h,k'}$ such that
\begin{equation} \label{eqn:err}
|E_{(k)} - E_{h,k'}| = O( d h)  
\end{equation}
 as $dh \rightarrow 0$, where $E_{(k)}$ is defined in (\ref{eqn:exc}). We will use the algorithms of Section \ref{sec:Algorithm} to approximate the low order eigenvalues of $M_h$, which as we have seen approximate the low order eigenvalues of $L$. For this, we need to construct the set of trial eigenvectors $\mathcal{S}$, and estimate its cardinality.
Recall that for the continuous operator, the set of candidate eigenvectors is derived using equation (\ref{hyp1}), and in particular by selecting the eigenvectors of $L^0$ that correspond to eigenvalues less or equal to $2M + B = 3M + E_{(s_j)}^0$. So for the discretized case we select the eigenvectors of $-\frac12 \Delta_h$ that correspond to eigenvalues less than or equal to $3M + E_{(s_j)}^0 + O(dh^2)$ due to equation (\ref{eq:cdh2}). Since $dh \rightarrow 0$, without loss of generality we slightly modify equation (\ref{hyp1}), to select the eigenvectors of $M_h$ that correspond to eigenvalues less than or equal to
\begin{equation} \label{eq:defB}
2M+ B = 3M + E_{(s_j)}^0 + 1
\end{equation}
 for sufficiently small  $h$, where this equation effectively redefines $B$ by increasing its value by $1$. Thus the cardinality of $\mathcal{S}$ in the case of the matrix $M_h$ follows from the continuous case and remains polynomial in $d$.

Specifically, we define
\begin{equation}   \label{eq:SB}
\mathcal{S}:= \{ u^0_{h,k} : E^0_{h,k} \leq 3M + E_{(s_j)}^0 + 1   \}
\end{equation}

\subsection{Algorithm for Excited State Energies}
\label{sec:AlgTISE} 
We now give the details of Algorithms  1 \& 2 of Section \ref{sec:Algorithm} applied to the time-independent Schr\"odinger equation (\ref{eqn:eqn}). Given $\e$, the algorithms produce the $j$ eigenvalue estimates $\tilde{E}_0 < ... < \tilde{E}_{j-1}$ of equation (\ref{eq:tildeE}).
Algorithm 1 computes $\tilde{E}_0$. For this, QPE \cite{NC} is applied repeatedly with its initial state taken to be every single element of the set of trial eigenvectors $\mathcal{S}$. We use  repetitions of the procedure to boost the success probability. 
We remark that our Algorithm 1 computes the ground state energy in a way similar to \cite{GS,Convex}, but under weakened assumptions.
Algorithm 2 iterates $j-1$ times the procedure of Algorithm 1, at each iteration producing the next estimate $\tilde{E}_i$ by taking into account all the previously produced estimates as we will explain below.

Both algorithms use QPE as the main module. The purpose is to compute approximations of the eigenvalues of the matrix $M_h$ of the previous section. Setting $N=2^{\lceil  2\log_2 (d/\e)  \rceil}$, we discretize (\ref{eqn:eqn}) with mesh size $h=\frac{1}{N+1} < \frac{\e^2}{d^2}$ to obtain the $N^d\times N^d$ matrix $M_h$, where we have $N^d = O((\frac{d}{\e})^{2d})$.  From (\ref{eqn:err}), we obtain that the low order matrix eigenvalues approximate the low order eigenvalues of the continuous operator with error proportional to $dh = O(\frac{\e^2}{d})$. 
The reason we have taken very small $h$ is because we want to ensure that $\e$-degenerate eigenvalues of the continuous operator will be approximated by tightly clustered eigenvalues of $M_h$. As $M$ is a constant, without loss of generality we may assume that $\e^{-1} \gg M$. Since the largest eigenvalue of $-\frac12\Delta_h$ is bounded from above by $2dh^{-2}$, and $V$ is uniformly bounded by $M$, we obtain that $\|M_h\|$ is bounded from above by $2dh^{-2}+M \ll 3dh^{-2}$, in the sense that $\frac{M}{dh^{-2}}=o(1)$. 

Let $R=3dh^{-2}$ and consider the matrix $W=e^{i M_h /R}$. Its eigenvalues are $e^{i E_h /R} = e^{ \frac{2\pi i E_h }{ 2 \pi R} } = e^{2\pi i \phi}$, where $E_h$ is an eigenvalue of $M_h$ and $\phi:= \frac{E_h}{2\pi R}$ denotes the corresponding phase.
 
QPE is used to compute an approximation $\hat{\phi}$ of $\phi$ with $b = 5\lceil \log_2 \frac{d}{\e} \rceil + 7$ bits of accuracy, and from this we get $\tilde{E} = 2\pi R\hat{\phi}$ so that 
\begin{equation}  \label{eq:totErr}
|E-\tilde{E}| \leq |E- E_h  | + |E_h - \tilde{E}| = O(\e) 
\end{equation}
where $E$ denotes the eigenvalue of $L$ that $E_h$ approximates according to (\ref{eqn:eqn}). QPE uses two registers, the top and the bottom. The size of the top register is related to the accuracy of QPE and its success probability. Recall that QPE succeeds when it produces an estimate with accuracy $2^{-b}$. The bottom register is used to hold an (approximate) eigenvector of $M_h$ corresponding to the phase of interest, and therefore has size $d\log_2 N = d\cdot O(\log \frac{d}{\e})$. The number of qubits in the top register is $t=b+t_0$, so that QPE has accuracy $2^{-b}$ with probability at least $1-\frac{1}{2(2^{t_0}-2)}$, assuming that an exact eigenvector is provided as initial state in the bottom register \cite[Sec. 5.2]{NC}. QPE uses powers of $W$, namely $W^{2^0}, W^{2^1},...,W^{2^t-1}$. We will approximate these powers using a splitting formula with error, as we will see below. This reduces the success probability of QPE to at least $p :=  1 - \frac{1}{2^{t_0} -2}$. We will set $t_0$ to be logarithmic in $d$, and will give all the details later on when dealing with the cost of our algorithm.

Consider an eigenvalue $E_h \leq B + 2M$ (see equations (\ref{eq:upperBound}) and (\ref{eq:defB})) of the matrix $M_h$ and let $u_h$ denote an eigenvector corresponding to $E_h$. Then QPE with initial state some $u^0_{h,i} \in \mathcal{S}$ succeeds with probability  at least $p_{u_h}(i):=| u_h^T u_{h,i}^0  |^2 \cdot p    = | u_h^T u_{h,i}^0  |^2    \cdot (1-\frac{1}{2^{t_0}-2})$  \cite{AbramsLloyd}.

Recall that $\mathcal{S}$ contains eigenvectors of $L^{0}$ that correspond to eigenvalues $E^0_h\leq B +2M$ as defined in (\ref{eq:SB}), and that the cardinality of $\mathcal{S}$ is polynomial in $d$. 
Applying the same approach of Section \ref{sec:PrelimAnal} for the eigenvectors of $M_h$, we conclude that 
for every eigenvector $u_h$ of the matrix $M_h$ that corresponds to an eigenvalue less than $B$, there exists a vector $u^0_{h,k} \in \mathcal{S}$ such that $| u_h^T u_{h,k}^0  |^2 \geq \frac{3}{4|\mathcal{S}|}$, where we have used equation (\ref{eqn:probBound}) with $c=2$ (since the value of $B$ we are using here leads to $c=2$ in this case too). 
Thus, after we run QPE with each element of $\mathcal{S}$ as initial state, the probability that at least one of the outcomes (in principle we do not know which one) will give a good estimate of  $E_{h}$ is at least $p_{u_h}(k) \geq \frac{3}{4|\mathcal{S}|}  p$.

We repeat the whole procedure $r$ times to boost the success probability of obtaining an estimate of $E_h$ with accuracy $\e$. Indeed, the probability that QPE fails with all initial states taken from $\mathcal{S}$ and in all its $r|\mathcal{S}|$ repetitions is
\begin{equation}
\left(  \prod_{i=1}^{|\mathcal{S}|} (1-p_{u_h}(i)) \right)^r \leq (1-p_{u_h}(k) )^r   \leq e^{-r p_{u_h}(k)} \leq e^{-r \frac{3}{4|\mathcal{S}|}p}
\end{equation}
Thus, the probability that at least one of the $r|\mathcal{S}|$ outcomes will lead to an approximation of $E_h$ with accuracy $O(\e)$ is at least 
\begin{equation}
1 -e^{-r \frac{3}{4|\mathcal{S}|}p}
= 1- e^{-r \frac{3}{4|\mathcal{S}|}   \cdot (1-\frac{1}{2^{t_0}-2})}
\end{equation}
We can boost this probability to be arbitrarily close to 1 by taking $r=poly(d)$, since $|\mathcal{S}|$ is polynomial in $d$.

Observe that Algorithm 1 selects the minimum measurement outcome from all the runs of QPE, and uses it to obtain $\tilde{E}_0$. Let this outcome be $m'\in\{ 0,\dots, 2^{t}-1\}$. The algorithm converts $m'$ to $m_0 =  \lfloor m' 2^{-t_0} \rfloor \in\{ 0,\dots, 2^{b}-1\}$ and uses it to obtain $\tilde{E}_0$, according to the formula 
\begin{equation}  \label{eq:Ephi}
\tilde{E}_0 = 2\pi R \hat{\phi}_0  = 2\pi R \frac{m_0}{2^b}.
\end{equation}
Since $|m'/2^t - \phi_0|\le 2^{-b}$ and $\phi_0, m'/2^t\in[ m_0/2^{b} , (m_0 +1 ) / 2^ b]  $, it follows that 
$ |2\pi R \phi_0 - \tilde{E}_0|  =      2\pi R \cdot |\phi_0 - m_0 / 2^ b| \leq 2\pi R / 2^b \leq \e$, which together with (\ref{eqn:err}) gives (\ref{eq:totErr}).

Let $G_0=\{ m : |\frac{m}{2^{b}}-\phi_0| \leq \frac{1}{2^b}  \}$.
Algorithm 1 fails either if none of the converted outcomes is an element of $G_0$, or at least one of the  converted outcomes is an element of $G_0$, but there is another converted outcome (produced by a failure of QPE) smaller than the minimum element of $G_0$. Thus, we can bound the total probability of failure by
\begin{eqnarray} \nonumber
\text{Pr(Algorithm 1 fails) } &=&\text{Pr(none of the outcomes leads to an element of } G_0\text{) } \\ \nonumber &+& \text{Pr(one of outcomes leads to an element of } G_0 \\
\nonumber &\ & \text{ but there is at least one other smaller converted} \\
\nonumber &\ & \text{ outcome)} \\
\nonumber &\le& e^{-r \frac{3}{4|\mathcal{S}|}p} \\
\nonumber &+& \text{Pr(QPE failed in at least one of} \\
\nonumber &\ & \text{the } r|\mathcal{S}|\text{ runs)} \\
\nonumber &\le& e^{-r \frac{3}{4|\mathcal{S}|}p} \\
\nonumber &+& (1 - \text{Pr(every run of QPE approximates }\\
\nonumber &\ & \text{one of the phase with error }  2^{-b}) ) \\
\nonumber &\le& e^{-r \frac{3}{4|\mathcal{S}|}p} + (1- p^{r|\mathcal{S}|}  ) \\
\label{eq:probBound}&\le& e^{-r \frac{3}{4|\mathcal{S}|}(1-\frac{1}{2^{t_0}-2})} + \left( 1- \left(1- \frac{1}{2^{t_0}-2}\right)^{r|\mathcal{S}|} \right)  \\
\nonumber &\le& e^{-r \frac{3}{4|\mathcal{S}|}(1-\frac{1}{2^{t_0}-2})} +  \frac{r|\mathcal{S}|}{2^{t_0}-2} , 
\end{eqnarray} 
where the third from last inequality follows from equation  (\ref{eq:probBound2}) below.
Observe that this bound can be made arbitrarily close to $0$ by selecting the number of repetitions $r$ to be a suitable polynomial in $d$, 
since $|\mathcal{S}|$ is polynomial in $d$, and by taking $t_0 = \beta \log d$, where $\beta$ is an appropriately chosen constant. We have used the fact that if a measurement outcome $\ell$ fails to estimate any of the phases, i.e. $|\frac{\ell}{2^b} - \phi_s| > 2^{-b}$ for all phases $\phi_s$ corresponding to eigenvalues of $M_h$, then 
\begin{equation}   \label{eq:probBound2}
\text{Pr}(\ell)=\sum_{s=0}^{N^d-1} c_s |\alpha(\ell,\phi_s)|^2 \leq \sum_{s=0}^{N^d-1} |c_s|^2 \frac{1}{2^{t_0}-2} = \frac{1}{2^{t_0}-2}     = (1- p).
\end{equation}
Here, the $c_s$ denote the projections of the initial state onto each of the eigenvectors of $M_h$, and
the $|\alpha(\ell,\phi_s)|^2$ denote the probability to get outcome $\ell$ given the exact eigenvector $u_{h,s}$ as input.
We have upper bounds for these quantities from \cite[Eq. 5.34]{NC}.
Therefore, the probability the measurement outcome estimates at least one (or, some) phase is $1-Pr(\ell) \ge p$. 

Recall that the set $\mathcal{S}$ has been constructed using an upper bound for $E_{(j-1)}$; see equations (\ref{eq:subseq}) and (\ref{eq:SB}). Algorithm 2 essentially repeats Algorithm 1 $(j-1)$ times, but selects the converted measurement outcome in a different way by considering the already selected outcomes.  At repetition $i$, it selects the minimum converted outcome $m_i$ that exceeds the outcome selected at the previous iteration by at least 2, i.e. $m_i \geq m_{i-1} +2$, where $m_i = \lfloor m'_i 2^{-t_0} \rfloor$ and $m'_i$ is a measurement outcome at the $i$th run, $i=1,2,3, j-1$; see also equation (\ref{eq:Ephi}).
The success probability for both Algorithm 1 and Algorithm 2 follows from (\ref{eq:probBound}) and is at least
\begin{equation} \label{eq:TotalSuccProb}
\left(1-\left(e^{-r \frac{3}{4|\mathcal{S}|}(1-\frac{1}{2^{t_0}-2})} +  \frac{r|\mathcal{S}|}{2^{t_0}-2} \right)\;\right)^j, 
\end{equation}
which can be made arbitrarily close to 1 by selecting $r$ to be a suitable polynomial in $d$ and taking $t_0$ to be sufficiently large.

Note that Algorithm 2 computes $\tilde{E}_i = 2\pi R \frac{m_i}{2^b}$, $i=1,2,\dots, j-1$, as estimates of the eigenvalues according to equation (\ref{eq:tildeE}) and the conditions \textbf{C1} and \textbf{C2} that follow it.
If both algorithms are successful with high probability, at the $i$th run we have that there exists a phase $\phi_i$ corresponding to an eigenvalue of $M_h$ such that $|2\pi R \phi_i - \tilde{E}_i| \leq \frac{2\pi R}{2^b} \leq  \e$. The condition $m_i \geq m_{i-1} +2$ in the selection of measurement outcomes guarantees that for any two  
$i_1\neq i_2$, the computed matrix eigenvalue approximations satisfy $\tilde{E}_{i_1}\neq \tilde{E}_{i_2}$ and 
$|\tilde{E}_{i_1} -\tilde{E}_{i_2}| = \Omega(\e)$
because for the corresponding phases we have $\phi_{i_1} \neq \phi_{i_2}$ as belonging to different intervals;
see Figure~\ref{Fig:phaseEstimation}. Moreover, the $\tilde E_{i_1}$ and $\tilde E_{i_2}$ also approximate different eigenvalues $E_{i_1}\ne E_{i_2}$ of the continuous operator because we have used a very fine discretization.
Finally, the algorithm does not fail to produce consecutive eigenvalues unless they differ by less than $O(\e)$ because we always select the minimum outcome that satisfies $m_i \geq m_{i-1} +2$.

\subsubsection{Cost of Quantum Phase Estimation}
\label{sec:Cost}

Algorithms 1 \& 2 use QPE as a module. The cost of QPE depends on the cost to prepare its initial state, and on the cost to implement the matrix exponentials $W^{2^0},W^{2^1},..W^{2^{t-1}}$, where $W= e^{i M_h/R}$. We approximate these exponentials below using Suzuki-Trotter splitting, the analysis of which proceeds similarly to that of \cite{PZ12,GS}

The initial states are taken from $\mathcal{S}$ which contains eigenvectors of $-\frac12 \Delta_h$ according to (\ref{eq:SB}). Each eigenvector can be prepared efficiently using the quantum Fourier transform, which diagonalizes the Laplacian, with a number of quantum operations proportional to $d \cdot \log^2 \frac{d}{\e}$ and using number of qubits $\log_2 N^d = d \cdot O(\log \frac{d}{\e})$. We remark that from the tensor product structure of the eigenvectors of $-\frac12 \Delta_h$, it suffices to prepare eigenvectors of the one-dimensional Laplacian; see e.g. \cite{Poisson,Klapp,Wavelet}.

Now let us turn to the approximation of the matrix exponentials. We simulate the evolution of the Hamiltonian $H = M_h/R$ for times $2^\tau$, 
 $\tau=0,1,\dots, t-1$, where we have set $t=b+t_0$. Let $H = H_1 + H_2$
where $H_1 = -\Delta_h / 2R $ and $H_2 = V_h/ R$, where we assume $V$ is given by an oracle.

To simulate quantum evolution by $H_1$, assuming the known eigenvalues of $-\frac12 \Delta_h$ are given by a quantum query oracle with $O(\log \frac{1}{\e})$ bits of accuracy, we again use the quantum Fourier transform to diagonalize $H_1$ with cost (i.e., a number of quantum operations) bounded by $d \cdot O(\log^2 \frac{d}{\e})$, and requiring  a number
of qubits proportional to $d \log \frac{d}{\e}$. Alternatively, if the eigenvalues of $-\frac12 \Delta_h$ are implemented explicitly (without an oracle) by the quantum algorithm, then the number of quantum operations required is a low order polynomial in $d$ and $\log_2 \frac{1}{\e}$, and so is the number of qubits \cite{Poisson}. For simplicity, we will not pursue this alternative here.  
The evolution of a system with Hamiltonian $H_2$ can be implemented using two quantum queries returning the values of $V$ at the grid points, and phase kickback. The queries are similar to those in Grover's algorithm \cite{NC} and the function evaluations of $V$ are truncated to $O(\log\frac 1\e)$ bits. 

We use a splitting formula $S_{2k}$ of order $2k+1$, $k\ge 1$, to
approximate $W^{2^t}=e^{i (H_1+H_2)2^t}$ by a product of the form
\begin{equation}\label{eq:r50}
\prod_{\ell=1}^{N_t} e^{i A_\ell z_\ell},
\end{equation}
where $A_\ell \in\{H_1,H_2\}$ and suitable $z_\ell$ that depends on $t$ and $k$.

The splitting formula $S_{2k}$ is due to Suzuki \cite{Suzuki90,Suzuki91}. It is used to approximate 
$e^{i(B+C)\Delta t}$, where $B$ and $C$ are Hermitian matrices. This formula is defined recursively by
\begin{eqnarray*}
S_2(B,C,\Delta t) &=& e^{iB\Delta t/2} e^{iC\Delta t} e^{iB\Delta t/2} \\
S_{2k}(B,C,\Delta t) &=& [S_{2k-2}(B,C, p_k\Delta t) ]^2 S_{2k-2}(B,C, (1-4p_k) \Delta t) \\
&& \hspace{10pc} \times [S_{2k-2}(B,C, p_k\Delta t) ]^2,
\end{eqnarray*}
where $p_k= (4 - 4^{1/(2k-1)})^{-1}$, $k=2,3,\dots$. 

Unfolding the recurrence above and combining it with \cite[Thm. 1]{PZ12} we obtain that the approximation of $W^{2^\tau}$ has the form
\begin{equation}
\label{eq:unfold}
\widetilde W^{2^\tau} = e^{iH_1 a_{\tau,0}} e^{iH_2 b_{\tau,1}} e^{i H_1 a_{\tau,1}} \cdots e^{iH_2 b_{\tau,L_\tau}} e^{i H_1 a_{\tau,L_\tau}},
\end{equation}
where
$a_{\tau,0},\dots, a_{\tau,L_\tau}$ and $b_{\tau,1},\dots,b_{\tau,L_\tau}$ and $L_\tau$ are parameters, $\tau=0,\dots, t_0+b-1$. 
The number of exponentials involving $H_1$ and $H_2$ in the expression above is $N_\tau=2L_\tau+1$. An explicit algorithm for computing each $\widetilde W^{2^\tau}$ is given in \cite{GS}.

Let $\| \cdot\|$ be the matrix norm induced by the Euclidean vector norm.
From \cite[Thm. 1 \& Cor. 1]{PZ12} the number $N_t$ of exponentials needed to approximate $W^{2^t}$
by a splitting formula of order $2k+1$ with error $\e_\tau$, $\tau=0,\dots, t_0+b-1$, is
\begin{equation*}
N_\tau \leq  16e \|H_1\| 2^\tau\, \left(\frac {25}3\right)^{k-1}
\left(\frac{8e\,2^\tau\|H_2\|}{\e_\tau}\right)^{1/(2k)}, 
\end{equation*}
for any $k \geq 1$. 
Since we want to approximate all the $W^{2^\tau}$, $\tau=0,1,...,t_0+b-1$, we sum the number of exponentials required to approximate each one of them. 
Thus the total number of matrix exponentials required by Algorithm 2, $\mathcal{N}_{tot}$, is bounded from above by
\begin{eqnarray}
\mathcal{N}_{tot}&=&jr|\mathcal{S}| \sum_{\tau=0}^{t_0+b-1} N_\tau \label{eq:numexp}  \nonumber \\
&\le&  jr|\mathcal{S}|  \left( 16e \| H_1\| \left(\frac {25}3\right)^{k-1}
 \left(8e\|H_2\|\right)^{1/(2k)}  \sum_{\tau=0}^{t_0+b-1}2^\tau
\left( \frac {2^\tau}{\e_\tau}\right)^{1/(2k)} \right). 
\end{eqnarray}
The factor $jr|\mathcal{S}|$ is the number of executions of QPE performed by our algorithms, and the second factor is the cost of a single QPE. Note that $j$ is the number of eigenvalues we wish to estimate, $|\mathcal{S}|$ is the number of eigenvectors we use as initial states, and $r$ is the number of times we repeat QPE per initial state to boost the success probability of getting the desired outcome.
We select a polynomial $g(d)$ such that the product $r|\mathcal{S}|/g(d) = o(1)$ (as $d\rightarrow \infty$). We then select the error of each exponential to be $\e_\tau = \tfrac {2^{\tau+1 - (b + t_0)}}{40 g(d)}$,
$\tau=0,\dots,t_0+b-1$. It is easy to check that $\sum_{\tau=0}^{t_0+b-1}\e_\tau \le \tfrac 1{20g(d)}$.
Thus the success probability of QPE is reduced by at most twice this amount \cite[p. 195]{NC}, giving $1- \frac{1}{2(2^{t_0}-2)} - \frac{1}{10g(d)}$. Next we set $t_0 = \lfloor \log_2 (5g(d) + 2)    \rfloor$, to get $p = 1 - \frac{1}{2^{t_0} -2}$ that we used above in deriving equation (\ref{eq:probBound}). Our choice of $g(d)$ and $t_0$ aims to make the bound of equation (\ref{eq:TotalSuccProb}) arbitrarily close to 1.

The largest eigenvalue of $-\Delta_h$ is $4dh^{-2} \sin^2(\pi N h/2) < 4dh^{-2}$. Since $R=3dh^{-2}$,  $H_1 = - \frac{1}{2}\Delta_h/ R = - \frac12\frac{1}{3dh^{-2}}\Delta_h$ and we have
$\norm{H_1} \leq \frac{2dh^{-2}}{3dh^{-2}} = \frac23$. 
Since $V$ is uniformly bounded by $M$ and $H_2 = V_h/ R$ we 
have $\norm{H_2} \leq M/3 d h^{-2}$. 
Substituting the value of $\e_\tau$ in (\ref{eq:numexp}), yields that the algorithm uses a number of
exponentials of $H_1$ and $H_2$ that satisfies
\begin{eqnarray*}
\mathcal{N}_{tot}
&\le&  jr|\mathcal{S}|  \left( 16e \| H_1\| \left(\frac {25}3\right)^{k-1}
 \left(8e\|H_2\|\right)^{1/(2k)}  \right) \sum_{\tau=0}^{t_0+b-1}2^\tau
\left( \frac {40 g(d) 2^\tau}{ 2^{\tau+1 - (b + t_0)}}\right)^{1/(2k)}  \\
&\leq&  jr|\mathcal{S}|  \left( 16e \| H_1\| \left(\frac {25}3\right)^{k-1}
 \left(8e\|H_2\|\right)^{1/(2k)}  \right)
\left( 20 g(d) 2^{t_0+b}\right)^{1/(2k)}  \\
&\leq&  jr|\mathcal{S}| \left( 16e \| H_1\| 2^{t_0 + b} \left(\frac {25}3\right)^{k-1}
\left( 160e\,2^{t_0 +b}\|H_2\| g(d) \right)^{1/(2k)} \right) .
\end{eqnarray*}
Using the bounds on $\|H_1\|$ and $\|H_2\|$, we obtain
$$
  \mathcal{N}_{tot} \leq  jr|\mathcal{S}| \left(  \frac{32e}{3}     2^{t_0 + b} \left(\frac {25}3\right)^{k-1}
\left( 160e\,2^{t_0 +b}   \frac{Mh^2}{3d}    g(d) \right)^{1/(2k)} \right).
$$
%
From $b = 5 \lceil \log_2 \tfrac{d}{\e} \rceil +7$, we have $2^b = 2^{5\lceil log_2 \frac{d}{\e}  \rceil  +7} \leq 2^{12} \left( \frac{d}{\e} \right)^5 = O(\frac{d^5}{\e^5})$.
Since $h < \frac{\e^2}{d^2}$, we have $2^b h^2 \leq 2^{12} \frac{d}{\e} =   O(\frac{d}{\e})$. 
Also, $2^{t_0} = 2^{\lfloor \log_2 (5g(d) + 2)    \rfloor} \leq 5g(d) +2$. 
We obtain
\begin{eqnarray} \label{eq:numexp2}
 \mathcal{N}_{tot} &\leq&  jr|\mathcal{S}| \left(  \frac{32e}{3}   \left(5g(d) + 2\right) \left( 2^{12} \left( \frac{d}{\e} \right)^5 \right)  \left(\frac {25}3\right)^{k-1} \right)\nonumber \\  
 &\qquad&  \qquad \times  
 \left( 160e\, \left(5g(d) + 2\right)  \left(  2^{12} \frac{d}{\e} \right)  \frac{M}{3d}    g(d) \right)^{1/(2k)} \nonumber \\
 &\leq&  jr|\mathcal{S}|\left( \widetilde C\; \frac{d^5 g(d)}{\e^5} \left(\frac {25}3\right)^{k-1} \left(\widehat{C}\;
\frac{g^2(d)}{\e}        \right)^{1/(2k)} \right) ,
%
\end{eqnarray}
for any $k>0$, where $\widetilde C$ and $\widehat{C}$ are suitable constants.

The {\it optimal} $k^*$, i.e., the one minimizing the upper bound for $\mathcal{N}_{tot} $
in (\ref{eq:numexp2}), 
is obtained in \cite[Sec. 5]{PZ12} and is given by
\[ k^* = \left\lfloor \sqrt{\frac{1}{2}  \log_{25/3}   \left(\widehat{C}\;
\frac{g^2(d)}{\e}        \right)     } + \frac 12 \right\rfloor = 
  \bar{C} \sqrt{\ln \frac {d}{\e}},
\]
for a suitable constant $\bar{C}$, 
since $g(d)$ is a polynomial in $d$ and we are taking its logarithm. 
With  $k^*$ and using again \cite[Sec. 5]{PZ12}, equation (\ref{eq:numexp2}) yields 
%
\begin{equation}\label{eq:numqueries}
\mathcal{N}_{tot}^{*} \leq \widetilde{C}   jr|\mathcal{S}| \frac{d^5}{\e^5}\; g(d)\;  e^{ 2\bar{C} \sqrt{ \ln \frac{25}{3}   \ln \frac{d}{\e}}}
=O \left(  g^2(d) \left(\frac{d}{\e}\right)^{5+\eta} \right) \quad {\rm as\ } d\e\to 0,
\end{equation}
where we have used $j=O(1)$ and $r|\mathcal{S}|=o(g(d))$, and where the equality above holds asymptotically for arbitrarily small $\eta>0$. 

We remark that of the $N_{tot}^*$ matrix exponentials roughly half involve $H_1$ and the remaining involve $H_2$; see (\ref{eq:unfold}). Since each exponential involving $H_2$ requires two queries the total number of queries is also of order $N_{tot}^*$. The cost to prepare the initial state, to diagonalize $-\frac12 \Delta_h$, and to implement the inverse Fourier transform that is applied prior to measurement in QPE, is proportional to 
$$ d\log^2 \frac{d}{\e} + (t_0+b)^2  =  O\left(d\log^2 \frac{d}{\e} \right),$$
since $t_0+b = O(\log \frac{d}{\e})$. 
Hence, the total number of quantum operations, excluding queries, is proportional to 
\begin{equation}\label{eq:numops}
\mathcal{N}_{tot}^* \cdot d\log^2 \frac{d}{\e}.
\end{equation}
Equations 
(\ref{eq:numqueries}) and (\ref{eq:numops}) yield 
that the total cost of the algorithm, including the number of
queries and the number of all other quantum operations, is proportional to
$$d\: g^2(d)  \left( \frac{d}{\e}\right)^{5 + \delta} ,$$
where $\delta >0$ is arbitrarily small. 
Finally, using equation (\ref{eq:TotalSuccProb}) we can select $r$ to be polynomial in $d$ and obtain success probability at least $\frac{3}{4}$, and the cost remains polynomial in $\frac{1}{\e}$ and $d$. We summarize our results in the following theorem. 
\vskip 2em
\begin{theorem} 
\label{thm1}
Consider the time-independent Schr\"odinger equation  (\ref{eqn:eqn})  on the $d$-dimensional unit cube with Dirichlet boundary conditions and where the potential $V$ and its first-order derivatives are uniformly bounded. 
Algorithms 1 \& 2 of Section \ref{sec:Algorithm} compute approximations of  $j=O(1)$ low order eigenvalues with error $O(\e)$, and satisfying conditions {\bf{C1}} and {\bf{C2}} of Section \ref{sec:ProbDef} with probability at least
$$\left(1-\left(e^{-r \frac{3}{4|\mathcal{S}|}(1-\frac{1}{2^{t_0}-2})} +  \frac{r|\mathcal{S}|}{2^{t_0}-2} \right)\;\right)^j, $$
where $r$ and $|\mathcal{S}|$ are polynomial in $d$, 
$t_0 = \lfloor \log_2 (5g(d) + 2)    \rfloor$, 
and $g(d)$ is a polynomial in $d$ selected such that $r|\mathcal{S}| = o(g(d))$.
The algorithms apply QPE with initial state each element of a set of trial eigenvectors $\mathcal{S}$, and repeat this
procedure $r$ times.

They use a number of queries proportional to
$$ \left( \frac{d}{\e} \right)^{5+\delta} \; g^2(d) 
\quad {\rm as\ } d\e\to 0, $$
a number of quantum operations excluding queries proportional to
$$\left( \frac{d}{\e} \right)^{5+\delta} \; d\;g^2(d)\;       \quad {\rm as\ } d\e\to 0,$$
where $\delta >0$ is arbitrarily small. The algorithms use a number of qubits proportional to
$$ d\; \log \frac{d}{\e} +  \log g(d).$$
\end{theorem}
\vskip 2em
\begin{rem}
The $5$ in the exponent of $\frac{d}{\e}$ is due to the fact that for simplicity we have taken $N=2^{\lceil2\log_2 \frac{d}{\e} \rceil}$ in the discretization of the continuous operator and our consequent choice of $b$, the number of bits of accuracy of QPE. As we explained, the purpose of the fine discretization is to ensure that degenerate eigenvalues of the continuous problem are approximated by tightly clustered eigenvalues of the matrix. It is possible to reduce the cost estimates of Theorem \ref{thm1} by taking $N=2^{\lceil (1+\gamma) \log_2 \frac{d}{\e} \rceil}$ and $b = (3+2\gamma)\lceil \log_2 \frac{d}{\e}\rceil +7$ where $\gamma \in (0,1)$ is a suitable constant. 
Then the total cost  becomes 
proportional to $d\;g^2(d)\;  \left( \frac{d}{\e}\right)^{3 + \delta + 2\gamma}$. 
 We do not pursue this any further since out goal was to establish
an algorithm with cost polynomial in $d$ and $\e^{-1}$.

On the other hand, the classical complexity of approximating a constant number of low order eigenvalues with error $\epsilon$ grows as $\left(\frac{1}{\epsilon}\right)^d$  in the deterministic worst case. Therefore, the problem suffers from the curse of dimensionality. 
The lower bound follows from the corresponding lower bounds for approximating the ground state energy with $\|V\| \leq 1$ , since more general conditions on $V$ are considered in this paper. 
A detailed discussion concerning the classical complexity lower bounds for approximating the ground state energy can be found in \cite{Qspeedup}. Since our quantum algorithm for this problem has cost polynomial in $d$ and $\frac{1}{\epsilon}$, it vanquishes the curse of dimensionality. 
\end{rem}

\section{Conclusion}
\label{Conclusion}

There are a number of recent results 
suggesting that certain eigenvalue problems are very hard, 
even for quantum computers. On the other hand, obtaining positive results for eigenvalue problems showing advantages of quantum computers over classical computers is particularly important. We discuss such a positive result for the approximation of the ground state energy in \cite{Qspeedup}, yet there is much more to be done.

In this paper, we consider the approximation of ground and excited state energies (low order eigenvalues) of a self-adjoint operator $L$. 
Typically, $L$ is discretized to yield a matrix eigenvalue problem. Since $L$ is self-adjoint, the resulting matrix is symmetric. It is important that the discretization is such that the eigenvalues of interest of $L$ are approximated accurately by matrix eigenvalues. This usually increases the matrix size. There are numerous classical algorithms that approximate matrix eigenvalues and/or the corresponding eigenvectors. In the case of symmetric matrices, we can approximate the eigenvalues that belong to a given range using the bisection method. 
In general, the cost of classical algorithms is bounded from below by the matrix size. Thus, the problem becomes hard when the matrix size is huge. On the other hand, quantum algorithms provide a way of overcoming this difficulty in certain cases. This can be accomplished using QPE as a module for approximating individual eigenvalues. 
QPE is efficient as long as two conditions are met. The first condition is that we are able to perform efficient quantum Hamiltonian simulation for the matrix whose eigenvalues are sought. There are numerous papers in the literature providing conditions and algorithms for efficient Hamiltonian simulation on a quantum computer. When a Hamiltonian is given by an oracle, efficient simulation can result from quantum parallelism and splitting formulas. The second condition needed in QPE is that we can prepare efficiently a quantum state encoding an approximation of an eigenvector corresponding to an eigenvalue of interest. 
It is enough that the approximate eigenvector is relatively good but by no means perfect. It suffices that the approximate eigenvector does not have an exponentially small (in the problem parameters) overlap with the unknown eigenvector. In our case, since we will be approximating a number of eigenvalues, we need an equal number of approximate eigenvectors. Obtaining the necessary approximate eigenvectors is a difficult task for arbitrary $L$.

We propose a way to overcome this difficulty and construct a set of approximate eigenvectors, which we call \textit{trial eigenvectors}, using the structure of $L$. Since $L=L^0+V$, and we have assumed that we know the eigenvalues and eigenvectors of $L^0$, we use a perturbation argument to construct the set of trial eigenvectors. We select eigenvectors of $L^0$ corresponding to a particular range of its eigenvalues. Equivalently, we exclude the eigenvectors of $L^0$ that correspond to eigenvalues which significantly exceed the eigenvalues of $L$ that we wish to approximate. We describe an algorithm that uses QPE, and the set of trial of eigenvectors, in order to approximate low order eigenvalues. If the cardinality of the set of trial eigenvectors is relatively small, i.e. its size is at most polynomial in the problem parameters, then our algorithm is efficient. 

In summary, general conditions for the approximation of ground and excited state energies on a quantum computer follow by combining conditions for efficient quantum simulation and for deriving a relatively small set of trial eigenvectors that can be implemented efficiently as quantum states. 

We illustrate how these general conditions are met 
in a special case of the time-independent Schr\"odinger equation with $d$ degrees of freedom. We show how our algorithm approximates a number of low order eigenvalues with error $\e$ and high probability. From our earlier work on this problem, we know that the complexity of approximating the ground state eigenvalue on a classical computer grows exponentially with the number of degrees of freedom in the worst-case.
Therefore, under the same or weaker assumptions as in this paper, the approximation of low order eigenvalues on a  classical computer satisfies the same lower bound. The problem suffers the curse of dimensionality in the classical worst case. 
For quantum algorithms, our previous approaches for computing the ground state energy require stronger conditions on V than those we consider here, and these approaches do not extend to computing excited state energies. We have developed an entirely new approach to approximate not only the ground state energy, but also excited state energies, with cost polynomial in $d$ and $\e^{-1}$. Our quantum algorithm vanquishes the curse of dimensionality. 

We remark on several open problems. 
We have assumed that $L=L^0 +V$, but such a partition need not be unique, and different partitions may result in algorithms with significantly different costs. It is possible, that with additional assumptions, one would be able to determine suitable partitions leading to fast algorithms. 
 Such a characterization is an open problem.  We have provided a condition for constructing a set of trial eigenvectors $\mathcal{S}$. Improving this condition to minimize the size of the resulting set $\mathcal{S}$ is another open problem. Finally, in our initial investigation of quantum algorithms for eigenvalues problems, we considered strong assumptions on $V$ and obtained efficient algorithms. Progressively, we have weakened these assumptions. It is important to continue working in this direction to extend the scope of our algorithm.

\section*{Acknowledgements}
The authors would like to thank Joseph F. Traub for useful comments and suggestions. This research has been supported in part by NSF/DMS.

\addcontentsline{toc}{section}{References} 
\bibliographystyle{plain} 

\bibliography{bib}   

\begin{thebibliography}{10}

\bibitem{AbramsLloyd}
D.~S. Abrams and S.~Lloyd.
\newblock Quantum algorithm providing exponential speed increase for finding
  eigenvalues and eigenvectors.
\newblock {\em Phys. Rev. Lett.}, 83:5162--5165, Dec 1999.

\bibitem{AharonovTa-Shma}
D.~Aharonov and A.~Ta-Shma.
\newblock Adiabatic quantum state generation and statistical zero knowledge.
\newblock In {\em Proceedings of the thirty-fifth annual ACM symposium on
  Theory of computing}, pages 20--29. ACM, 2003.

\bibitem{andrews2011proof}
B.~Andrews and J.~Clutterbuck.
\newblock Proof of the fundamental gap conjecture.
\newblock {\em Journal of the American Mathematical Society}, 24(3):899--916,
  2011.

\bibitem{Aspuru-Guzik:2005yq}
A.~Aspuru-Guzik, A.~D. Dutoi, P.~J. Love, and M.~Head-Gordon.
\newblock Simulated quantum computation of molecular energies.
\newblock {\em Science}, 309(5741):1704--1707, 2005.

\bibitem{babuskaOsborn}
I.~Babuska and J.~Osborn.
\newblock Eigenvalue problems.
\newblock {\em Handbook of numerical analysis.}, 2:641, 1991.

\bibitem{Berry07}
D.~W. Berry, G.~Ahokas, R.~Cleve, and B.~C. Sanders.
\newblock Efficient quantum algorithms for simulating sparse hamiltonians.
\newblock {\em Communications in Mathematical Physics}, 270(2):359--371, 2007.

\bibitem{stoquastic}
S.~Bravyi, D.~P. Divincenzo., R.~I. Oliveira, and B.~M. Terhal.
\newblock The complexity of stoquastic local {H}amiltonian problems.
\newblock {\em Quantum Information and Computation}, 8(5):361--385, 2008.

\bibitem{Poisson}
Y.~Cao, A.~Papageorgiou, I.~Petras, J.~F. Traub, and S.~Kais.
\newblock Quantum algorithm and circuit design solving the {P}oisson equation.
\newblock {\em New Journal of Physics}, 15:013021, 2013.

\bibitem{childs2013bose}
A.~M. Childs, D.~Gosset, and Z.~Webb.
\newblock The bose-hubbard model is qma-complete.
\newblock In J.~Esparza, P.~Fraigniaud, T.~Husfeldt, and E.~Koutsoupias,
  editors, {\em Automata, Languages, and Programming}, volume 8572 of {\em
  Lecture Notes in Computer Science}, pages 308--319. Springer Berlin
  Heidelberg, 2014.

\bibitem{cullum2002lanczos}
J.~K. Cullum and R.~A. Willoughby.
\newblock {\em Lanczos Algorithms for Large Symmetric Eigenvalue Computations:
  Vol. 1: Theory}, volume~41.
\newblock SIAM, 2002.

\bibitem{Demmel}
J.~W. Demmel.
\newblock {\em Applied Numerical Linear Algebra}.
\newblock SIAM, Philadelphia, PA, 1997.

\bibitem{Feynman}
R.~Feynman.
\newblock Simulating physics with computers.
\newblock {\em SIAM Journal on Computing}, 26:1484--1509, 1982.

\bibitem{FW}
G.~E. Forsythe and W.~R. Wasow.
\newblock {\em Finite-{D}ifference {M}ethods for {P}artial {D}ifferential
  {E}quations}.
\newblock Dover, New York, 2004.

\bibitem{Oliv05}
F.~Furche and D.~Rappoport.
\newblock Density functional methods for excited states: equilibrium structure
  and electronic spectra.
\newblock In M.~Olivucci, editor, {\em Computational Photochemistry}, volume~16
  of {\em Theoretical and Computational Chemistry}, pages 93--128, Amsterdam,
  2005. Elsevier.

\bibitem{Golub}
G.~H. Golub and C.~F. Van~Loan.
\newblock {\em Matrix computations}.
\newblock JHU Press, 2012.

\bibitem{Gustafson}
Stephen~J Gustafson and Israel~Michael Sigal.
\newblock {\em Mathematical concepts of quantum mechanics}.
\newblock Universitext. Springer, 2011.

\bibitem{Sigal}
P.~D. Hislop and I.~M. Sigal.
\newblock {\em Introduction to Spectral Theory: With Applications to
  {S}chr{\"o}dinger Operators}.
\newblock Number v. 113 in Applied Mathematical Sciences Series. Springer
  Verlag, New York, 1996.

\bibitem{HK1}
P.~Hohenberg and W.~Kohn.
\newblock Inhomogeneous electron gas.
\newblock {\em Phys. Rev.}, 136(3B):B864, 1964.

\bibitem{Kassal}
I.~Kassal, J.~D. Whitfield, A.~Perdomo-Ortiz, M.-H. Yung, and A.Aspuru-Guzik.
\newblock Simulating chemistry using quantum computers.
\newblock {\em Ann. Rev. Phys. Chem.}, 62:185--207, 2011.

\bibitem{KempeLocalHam}
J.~Kempe, A.~Kitaev, and O.~Regev.
\newblock The complexity of the local {H}amiltonian problem.
\newblock {\em SIAM J. Comput.}, 35(5):1070--1097, 2006.

\bibitem{Klapp}
A.~Klappenecker and M.~R\"otteler.
\newblock Discrete cosine transforms on quantum computers.
\newblock In {\em Proceedings of the 2nd International Symposium on Image and
  Signal Processing and Analysis}, pages 464--468, 2001.

\bibitem{Lanyon10}
B.~P. Lanyon, J.~D. Whitfield, G.~G. Gillett, M.~E. Goggin, M.~P. Almeida,
  I.~Kassal, J.~D. Biamonte, M.~Mohseni, B.~J. Powell, M.~Barbieri,
  A.~Aspuru-Guzik, and A.~G. White.
\newblock Towards quantum chemistry on a quantum computer.
\newblock {\em Nature Chemistry}, 2:106--111, 2010.

\bibitem{Leveque}
R.~J. Leveque.
\newblock {\em Finite Difference Methods for Ordinary and Partial Differential
  Equations}.
\newblock SIAM, Philadelphia, PA., 2007.

\bibitem{Lloyd96}
S.~Lloyd.
\newblock Universal quantum simulators.
\newblock {\em Science 23}, 273(5278):1073--1078, 1996.

\bibitem{love2012back}
P.~J. Love.
\newblock Back to the future: A roadmap for quantum simulation from vintage
  quantum chemistry.
\newblock In S.~Kais, editor, {\em Quantum Information and Computation for
  Chemistry}, volume 154 of {\em Advances in Chemical Physics}, pages 39--66,
  Hoboken, NJ, 2014. Wiley.

\bibitem{NC}
M.~Nielsen and I.~Chuang.
\newblock {\em Quantum {C}omputation and {Q}uantum {I}nformation}.
\newblock Cambridge University Press, Cambridge UK, 2000.

\bibitem{AP07}
A.~Papageorgiou.
\newblock On the complexity of the multivariate {S}turm--{L}iouville eigenvalue
  problem.
\newblock {\em J. Complexity}, 23(4-6):802--827, 2007.

\bibitem{Convex}
A.~Papageorgiou and I.~Petras.
\newblock Estimating the ground state energy of the {S}chr{\"o}dinger equation
  for convex potentials.
\newblock {\em J. Complexity}, 30:469--494, 2014.

\bibitem{GS}
A.~Papageorgiou, I.~Petras, J.~F. Traub, and C.~Zhang.
\newblock A fast algorithm for approximating the ground state energy on a
  quantum computer.
\newblock {\em Mathematics of Computation}, 82(284):2293--2304, 2014.

\bibitem{Qspeedup}
A.~Papageorgiou and J.~F. Traub.
\newblock Measures of quantum computing speedup.
\newblock {\em Phys. Rev. A}, 88(2):022316, 2013.

\bibitem{PZ12}
A.~Papageorgiou and C.~Zhang.
\newblock On the efficiency of quantum algorithms for {H}amiltonian simulation.
\newblock {\em Quantum Information Processing}, 11:541--561, 2012.

\bibitem{Parlett}
B.~N. Parlett.
\newblock {\em The symmetric eigenvalue problem}, volume~7.
\newblock SIAM, 1980.

\bibitem{schuchDFT}
N.~Schuch and F.~Verstraete.
\newblock Computational complexity of interacting electrons and fundamental
  limitations of density functional theory.
\newblock {\em Nature Physics}, 5(10):732--735, 2009.

\bibitem{strangFiniteElement}
G.~Strang and G.~J. Fix.
\newblock {\em An analysis of the finite element method}.
\newblock Wellesley-Cambridge Press, 2 edition, 2008.

\bibitem{Suzuki90}
M.~Suzuki.
\newblock Fractal decomposition of exponential operators with applications to
  many-body theories and {M}onte {C}arlo simulations.
\newblock {\em Phys. Letters A}, 146(6):319--323, 1990.

\bibitem{Suzuki91}
M.~Suzuki.
\newblock General theory of fractal path integrals with application to
  many-body theories and statistical physics.
\newblock {\em J. Math. Phys.}, 32:400--407, 1991.

\bibitem{Titchmarsh}
E.C. Titchmarsh.
\newblock {\em Eigenfunction Expansions: Associated with Second-order
  Differential Equations}.
\newblock Eigenfunction Expansions Associated with Second-order Differential
  Equations. Oxford University Press, Oxford, UK, 1962.

\bibitem{intBosons}
T.-C. Wei, M.~Mosca, and A.~Nayak.
\newblock Interacting boson problems can be {Q}{M}{A} hard.
\newblock {\em Phys. Rev. Letters}, 104(4):040501, 2010.

\bibitem{Weinberger1}
H.~F. Weinberger.
\newblock Upper and lower bounds for eigenvalues by finite difference methods.
\newblock {\em Communications on Pure and Applied Mathematics}, 9(3):613--623,
  1956.

\bibitem{Weinberger2}
H.~F. Weinberger.
\newblock Lower bounds for higher eigenvalues by finite difference methods.
\newblock {\em Pacific J. Math}, 8(2):339--368, 1958.

\bibitem{Wavelet}
M.~V. Wickerhauser.
\newblock {\em Adapted wavelet analysis from theory to software}.
\newblock A.K. Peters, Wellesley, MA, 1994.

\bibitem{Zalka2}
C.~Zalka.
\newblock Efficient simulation of quantum systems by quantum computers.
\newblock {\em Fortschritte der Physik}, 46(6-8):877--879, 1998.

\bibitem{Zalka1}
C.~Zalka.
\newblock Simulating quantum systems on a quantum computer.
\newblock {\em Proceedings of the Royal Society of London. Series A:
  Mathematical, Physical and Engineering Sciences}, 454(1969):313--322, 1998.

\end{thebibliography}

%
%

\end{document}